# EIGENRAYS IN 3D HETEROGENEOUS ANISOTROPIC MEDIA: PART II – KINEMATICS, VALIDATION OF THE LAGRANGIAN

*Zvi Koren and Igor Ravve (corresponding author), Emerson*

[zvi.koren@emerson.com](mailto:zvi.koren@emerson.com) , [igor.ravve@emerson.com](mailto:igor.ravve@emerson.com)

## ABSTRACT

The form of the Lagrangian proposed in Part I of this study has been previously used for obtaining stationary ray paths between two endpoints in isotropic media. We extended it to general anisotropy by replacing the isotropic medium velocity with the ray (group) velocity magnitude which depends on both, the elastic properties at the ray location and the ray direction. This generalization for general anisotropy is not trivial and in this part we further elaborate on the correctness, physical interpretation, and advantages of this original arclength-related Lagrangian.

We also study alternative known Lagrangian forms and their relation to the proposed one. We then show that our proposed first-degree homogeneous Lagrangian (with respect to the ray direction vector) leads to the same kinematic ray equations as the alternative Lagrangians representing first- and second-degree homogeneous functions. Using different anisotropic examples, we further validate/demonstrate the correctness of the proposed Lagrangian, analytically (for a canonical case of an ellipsoidal orthorhombic medium) and numerically (including the most general medium scenario: spatially varying triclinic continua).

Finally, we analyze the commonly accepted statement that the Hamiltonian and the Lagrangian can be related via a resolvable Legendre transform only if the Lagrangian is a time-related homogeneous function of the second-degree with respect to the vector tangent to the ray. We show that this condition can be bypassed, and a first-degree homogeneous Lagrangian, with a singular Hessian matrix, can be used as well, when adding a fundamental physical constraint which turns to be the Legendre transform itself. In particular, the momentum equation can be solved, establishing, for example, the ray direction, given the slowness vector.

## INTRODUCTION

The Eigenray kinematic method can be considered a special type of the ray bending approach. An extensive review of the existing studies on the ray bending methods applied to the two-point kinematic ray tracing problem in isotropic and general anisotropic media has been listed in Part I. In this part (Part II), we concentrate on the validation of our proposed Lagrangian for general anisotropic media, its physical interpretation and its advantages for the implementation for obtaining the stationary ray paths using our proposed finite element solver (Part III of this study).

Choosing the proper Lagrangian applied for both, the kinematic ray tracing (KRT) and dynamic ray tracing (DRT), and the solution method to be used, in particular, for general anisotropic media, is a crucial step for the success of this challenging problem. At this time, we are not yet aware of any published complete formulation and application of the ray bending method for 3D heterogeneous general anisotropic media. We realize that (due to its originality) it is our duty to prove and validate the correctness of the proposed Lagrangian and to justify the theoretical aspects of its implementation. Indeed, this is not an easy task as it requires massive derivations and using examples involving different anisotropic scenarios. The readers interested only in the implementation of the Eigenray method can skip this part and move to Part III. However, we do believe that this part provides some deep physical insights in this challenging topic.

In Part I we emphasize that among the different alternative Lagrangians, we find the proposed arclength-related Lagrangian most physically plausible, and most convenient and effective for our proposed finite-element kinematic (Part III) and dynamic (Part VII) solutions. By physically plausible we mean that the ray (group) velocity inherent within the proposed Lagrangian complies with the physical nature of its directional dependency in anisotropic elastic continua (which is not always the case for some alternative Lagrangians; although, mathematically they can still yield correct kinematic equations). This topic will be later explained in details.

We recall that the form of our proposed Lagrangian has been previously used for solving ray bending problems in isotropic media, with the medium velocity $v(\mathbf{x})$ in the denominator (e.g., Červený 2002a, 2002b, for seismic waves; Holm, 2011, 2012; Chaves, 2016, in the context of electromagnetic waves) and it has been extensively tested. The extension to anisotropic media involves replacing the isotropic velocity $v(\mathbf{x})$ by the ray (group) velocity $v_{\text{ray}}(\mathbf{x},\mathbf{r})$, where $\mathbf{r}$ is the ray direction vector. This is not an easy task. The kinematics and dynamics of the Eigenray method for general anisotropy explicitly involve the spatial, directional and mixed derivatives of the Lagrangian, and hence also those of the ray velocity. In particular, the directional derivatives of the ray velocity should be properly (physically) implemented. These derivatives have been recently developed by Ravve and Koren (2019) and briefly summarized in Appendix E of Part I.

Červený (2002a, 2002b) published a fundamental theoretical work on Fermat's variational principle for anisotropic inhomogeneous media, where he stated that the Hamiltonian and the Lagrangian can be related via a resolvable Legendre transformation only if the Lagrangian is a homogeneous function of the second degree with respect to (wrt) the components of the ray velocity, $d\mathbf{x}/d\tau$, where the current traveltime $\tau$ is the flow parameter. "Resolvable" means a possibility to find the ray direction vector (for the flow variable arclength) or the ray velocity vector (for the flow variable traveltime) components from the momentum equation, given the elastic properties and the slowness vector at a given location. In this part, we explicitly refer to these important papers. We show that first-degree homogeneous Lagrangians can be also used, by adding an independent scalar physical constraint, which turns to be the Legendre transform itself. We provide the general form of this constraint for general Lagrangians with different flow parameters (e.g., time, arclength, sigma). We demonstrate analytically that our proposed, first-degree homogeneity, arclength-related Lagrangian and the two traveltime-related Lagrangians suggested in the cited works by Červený lead to the same kinematic equations.

In Part I, we provided the arclength-related Hamiltonian, $H(\mathbf{x},\mathbf{p})$, where $\mathbf{p}$ is the slowness vector. In this part (Part II), we show the connection of this Hamiltonian to the proposed Lagrangian, $L(\mathbf{x},\mathbf{r})$, through the Legendre transform. Next, using different anisotropic symmetries, we demonstrate analytically and numerically, that the kinematic relationships obtained from the proposed Lagrangian $L(\mathbf{x},\mathbf{r})$, its matching Hamiltonian $H(\mathbf{x},\mathbf{p})$, and an alternative "eigenvalue" Hamiltonian, $H^{\lambda}(\mathbf{x},\mathbf{p})$, suggested by Červený (2000, 2002a, 2002b) (see Table 1 of Part I), are all identical. For this, we provide the Hamiltonian ray tracing equations for both $H$ (in Part I) and $H^{\lambda}$ (in this Part II). The two Lagrangians suggested by Červený are based on the regular (invertible) Finsler metric tensor $\mathbf{G}$, and we discuss different ways to establish this matrix, based on the eigenvalue and the arclength-related Hamiltonians.

Appendices

The mathematical derivations and the numerical examples have been moved to the appendices.

In Appendix A we review and discuss the so-called unmodified first-degree and the modified second-degree time-related homogeneous Lagrangians suggested by Červený (2002a, 2002b), and the Hamiltonian matching the modified Lagrangian, which we call "the eigenvalue Hamiltonian".

In Appendix B we review the kinematic equations following from the (time-related) eigenvalue Hamiltonian. We arrange these equations in a tensor form (rather than a component-wise form widely used in the literature). The resulting expression of this appendix is the arclength derivative of the slowness vector, $d\mathbf{p}/ds$, that follows from the time-related eigenvalue Hamiltonian; this derivative is compared in the other appendices, numerically and analytically, with that obtained from the proposed arclength-related Lagrangian and arclength-related Hamiltonian. We then explicitly validate their equivalency for different anisotropic symmetries (see Appendices E, F and G).

In Appendix C we mainly discuss the Finsler metric tensor which is the kernel of the two Lagrangians suggested by Červený (2002a, 2002b); it also relates the slowness vector to the ray velocity vector. We present two ways to construct this tensor: with the time-related (polarization-dependent) eigenvalue Hamiltonian mentioned in the two cited works, and with an alternative time-related Hamiltonian, independent of the polarization vector. Both ways lead to identical results, but we find the latter (using the arclength-related Hamiltonian) more convenient and straightforward. We then discuss the use of the (well-known) polarization-dependent tensor (similar to the Christoffel tensor, where the slowness vector is replaced by the polarization vector), as an alternative to the Finsler metric. This tensor relates the slowness and the ray velocity vectors as well, and we study the inversibility of this tensor. However, we note that unlike the Finsler metric, this tensor can't be used to define the Lagrangian.

In Appendix D, we prove that although our proposed Lagrangian has a different form than those suggested by Červený, it yields the same kinematic equations. This proof can be considered another validation for the correctness of the proposed Lagrangian. The validation is then continued analytically for an elliptic medium and numerically for tilted orthorhombic and triclinic media in the other appendices.

Appendices E and F can be considered numerical examples. Considering a factorized anisotropic inhomogeneous (FAI) non-ellipsoidal orthorhombic medium (Appendix E) and a general spatially-varying triclinic medium (Appendix F), we validate the proposed Lagrangian numerically by computing the kinematic characteristics $\mathbf{r} = d\mathbf{x}/ds$ and $d\mathbf{p}/ds$ using the proposed Lagrangian, $L(s)$, and alternatively by using two different Hamiltonians, $H(s)$ and $H^\lambda(\tau)$, based on the Christoffel equation and on its root $\lambda$, respectively. We show that all methods yield identical results.

In Appendix G, we consider an ellipsoidal orthorhombic medium and we perform analytically the same validation process, as in Appendices E and F.

In Appendix H, we discuss the Hamiltonian- and Lagrangian-based approaches to compute the ray velocity direction $\mathbf{r}$, given the slowness vector $\mathbf{p}$ and the medium elastic properties. The solution of the Hamiltonian-based equation, $\mathbf{r} = H_\mathbf{p}(\mathbf{x},\mathbf{p})$, is straightforward and this is our actual approach in this study. The Lagrangian-based momentum equation, $L_\mathbf{r}(\mathbf{x},\mathbf{r}) = \mathbf{p}$, is a nonlinear (inversion) equation that can be solved numerically (iteratively). We show that although the resolving matrix (the Hessian of our first-degree homogeneous Lagrangian wrt the ray direction, $L_{\mathbf{rr}}$) is a positive semidefinite (noninvertible) matrix, the problem is resolvable with an additional scalar constraint enforcing the ray direction $\mathbf{r}$ to be normalized. We then provide the general constraint equation related to the alternative first-degree homogeneity Lagrangians corresponding to the different flow parameters (e.g., traveltime, arclength, sigma).

## THE ARCLENGTH-RELATED LAGRANGIAN

In Part I (equation 2) we proposed the arclength-related Lagrangian,

$$L(\mathbf{x},\mathbf{r}) = \frac{\sqrt{\mathbf{r} \cdot \mathbf{r}}}{v_{\text{ray}}(\mathbf{x},\mathbf{r})} \quad , \tag{1}$$

where $\mathbf{x}$ is the location vector, $\mathbf{r} \equiv \dot{\mathbf{x}} \equiv d\mathbf{x}/ds$ is the normalized ray direction (or ray velocity direction) vector, and $s$ is the arclength flow parameter. However, we did not explain where the numerator $\sqrt{\mathbf{r} \cdot \mathbf{r}}$ comes from. In this part we will elaborate on the proposed Lagrangian, its correctness for general anisotropy and its relations to the other known Lagrangians.

Let us start with a 2D case, considering a general anisotropic medium and a stationary or approximated ray, given by a unique function, $y = y(x)$. The elementary traveltime $d\tau$ can be written as,

$$d\tau = \frac{ds}{v_{\text{ray}}(x, y, y')} = \frac{\sqrt{dx^2 + dy^2}}{v_{\text{ray}}(x, y, y')} = \frac{\sqrt{1+(dy/dx)^2}}{v_{\text{ray}}(x, y, y')}dx$$
$$= \frac{\sqrt{1+y'^2}\,dx}{v_{\text{ray}}(x, y, y')} = L(x, y, y')dx \quad . \tag{2}$$

The derivative $y'(x)$ defines the slope of the 2D trajectory and thus affects the direction-dependent anisotropic ray velocity. The Lagrangian of equation 2 is the traveltime integrand,

$$L = \frac{d\tau}{dx} = \frac{\sqrt{1+y'^2}}{v_{\text{ray}}(x, y, y')} \quad , \quad t = \int_S^R d\tau = \int_S^R \frac{d\tau}{dx}dx = \int_S^R L(x, y, y')dx \quad . \tag{3}$$

Note that the Lagrangian in equation 3 is not a homogeneous function (of whatever degree) wrt the slope, $y'(x)$. Furthermore, both, its numerator and denominator are not homogeneous functions. Considering the numerator, $\sqrt{1+(k\,y')^2} \neq k^n \sqrt{1+y'^2}$ (where $n$ is fixed and $k$ is arbitrary). Considering the denominator, the positive factor $k$ in $v_{\text{ray}}(x, y, k\,y')$ means the ray velocity magnitude for a quite different direction, $\arctan(k\,y')$ instead of $\arctan y'$ (measured from the horizontal axis). Obviously, this change of direction results in the velocity that does not comply a homogeneous function.

However, even in the 2D case, $y$ is not necessarily a unique function of $x$, and it becomes more suitable to consider both position components, $x$ and $y$, as functions of a single monotonously increasing parameter (a flow parameter), such as the current traveltime $\tau$, or the arclength $s$. We choose the arclength, and equation 1 becomes,

$$L = \frac{d\tau}{ds} = \frac{\sqrt{(dx/ds)^2 + (dy/dx)^2}\,ds}{v_{\text{ray}}(x,y,dx/ds,dy/ds)} = \frac{\sqrt{\dot{x}^2 + \dot{y}^2}\,ds}{v_{\text{ray}}(x,y,\dot{x},\dot{y})} \quad , \tag{4}$$

where we recall that dot over a symbol means its derivative WRT the arclength, $s$. This relationship can be naturally expanded to the 3D case,

$$L(s) = \frac{d\tau}{ds} = \frac{\sqrt{\dot{x}_1^2 + \dot{x}_2^2 + \dot{x}_3^2}}{v_{\text{ray}}(x_1,x_2,x_3,\dot{x}_1,\dot{x}_2,\dot{x}_3)} = \frac{\sqrt{\dot{\mathbf{x}} \cdot \dot{\mathbf{x}}}}{v_{\text{ray}}(\mathbf{x},\mathbf{r})} = \frac{\sqrt{\mathbf{r} \cdot \mathbf{r}}}{v_{\text{ray}}(\mathbf{x},\mathbf{r})} \quad , \quad t = \int_A^B L(s)\,ds \quad . \tag{5}$$

Of course, since $\mathbf{r} \cdot \mathbf{r} = 1$, if our only goal is to compute the stationary or non-stationary time along the ray path or its approximation, respectively, setting the numerator to be 1 is legitimate and one can ignore the factor $\sqrt{\mathbf{r} \cdot \mathbf{r}}$. However, our task is to obtain a stationary ray with the Fermat's principle,

$$t = \int_A^B L(s)\,ds \to \text{stationary} \quad , \tag{6}$$

and for that we need to compute the spatial and directional gradients of the Lagrangian for the kinematic ray tracing (KRT), and its spatial, directional and mixed Hessians for the dynamic ray tracing (DRT). This makes the abovementioned substitution illegal as it leads to the loss of the most important terms; thus, $\sqrt{\mathbf{r} \cdot \mathbf{r}}$ is a necessary factor in the numerator of the proposed Lagrangian.

The proposed Lagrangian is of first-degree homogeneity wrt the ray direction $\mathbf{r}$. This is true for the 2D case in equation 4 and for the 3D case in equation 5; we will further consider the general 3D case only. In order to establish the homogeneity degree of the Lagrangian, consider its numerator and denominator separately. If a positive scaling factor $k$, if applied to the ray direction in the numerator of the Lagrangian, it affects the value indicating that the numerator is the first-degree homogeneous function wrt the tangent vector $k\mathbf{r}$ (non-normalized ray direction),

$$\sqrt{k\mathbf{r}\cdot k\mathbf{r}} = k\sqrt{\mathbf{r}\cdot\mathbf{r}} \qquad . \qquad (7)$$

Now, consider the ray velocity magnitude in the denominator of the Lagrangian,

$$v_{\text{ray}}(\mathbf{x}, k\mathbf{r}) = v_{\text{ray}}(\mathbf{x}, \mathbf{r}) \quad , \quad \partial v_{\text{ray}} / \partial k = 0 \qquad , \qquad (8)$$

where $k\mathbf{r} = \dot{\mathbf{x}}_\zeta = d\mathbf{x}/d\zeta$ is a vector, tangent to the ray and not necessarily normalized, and $\zeta$ is the corresponding generic (arbitrary) flow variable; the scale factor may depend on this variable, $k = k(\zeta)$. Equation 8 emphasizes that the ray velocity magnitude is a physical parameter (unique for compressional waves and non-unique for shear waves) independent of the direction scale, $k$. This means that $v_{\text{ray}}(\mathbf{x}, \mathbf{r})$ is a zero-degree homogeneous function wrt the components of the ray direction vector. This magnitude is defined by the medium properties $\tilde{\mathbf{C}}$ at the given location $\mathbf{x}$, and the ray direction $\mathbf{r}$. The direction represents a vector tangent to the ray, and the length of this tangent vector is inessential for computing the ray velocity magnitude. Moreover, if one chooses to use a generic flow variable, $\zeta$, and the corresponding vector tangent to the ray, $\dot{\mathbf{x}}_\zeta = d\mathbf{x}/d\zeta$ (which is not necessarily normalized), then the magnitude of the ray velocity should be formulated for the normalized tangent vector,

$$\mathbf{r} = \frac{\dot{\mathbf{x}}_\zeta}{\sqrt{\dot{\mathbf{x}}_\zeta \cdot \dot{\mathbf{x}}_\zeta}} \quad , \quad v_{\text{ray}} = v_{\text{ray}}(\mathbf{x}, \mathbf{r}) = v_{\text{ray}}\left(\mathbf{x}, \frac{\dot{\mathbf{x}}_\zeta}{\sqrt{\dot{\mathbf{x}}_\zeta \cdot \dot{\mathbf{x}}_\zeta}}\right) \qquad , \qquad (9)$$

which also follows from equations 18 and 19 of Part I.

Next, we analyze these two equations (18 and 19 of Part I) for the homogeneity degree of the slowness vector $\mathbf{p}$ and the ray velocity magnitude $v_{\text{ray}}$ wrt the tangent vector. Given the ray direction (or ray velocity direction), the ray velocity magnitude is established in two stages. First we compute the slowness vector, $\mathbf{p}$, applying the collinearity of the (reference) Hamiltonian gradient wrt the slowness vector, $H_{\mathbf{p}}^{\bar{\tau}}$, and the ray direction, $\mathbf{r}$,

$$H_{\mathbf{p}}^{\bar{\tau}}(\mathbf{x},\mathbf{p}) \times \mathbf{r} = 0 \quad , \quad H^{\bar{\tau}}(\mathbf{x},\mathbf{p}) = 0 \quad . \quad (10)$$

Replacing $\mathbf{r}$ by $k\,\mathbf{r}$ has no effect on the solution of equation 10, which demonstrates that the slowness vector is a zero-degree homogeneous function wrt the ray direction,

$$\mathbf{p} = \mathbf{p}(\mathbf{x},\mathbf{r}) = \mathbf{p}\left(\mathbf{x}, \frac{\dot{\mathbf{x}}_\zeta}{\sqrt{\dot{\mathbf{x}}_\zeta \cdot \dot{\mathbf{x}}_\zeta}}\right) \quad . \quad (11)$$

At the second stage, we compute the ray velocity magnitude, applying the well-known identity, $\mathbf{p} \cdot \mathbf{v}_{\text{ray}} = 1$, leading to,

$$\frac{1}{v_{\text{ray}}} = \mathbf{p} \cdot \mathbf{r} = \frac{\mathbf{p} \cdot \dot{\mathbf{x}}_\zeta}{\sqrt{\dot{\mathbf{x}}_\zeta \cdot \dot{\mathbf{x}}_\zeta}} \quad , \quad (12)$$

which demonstrates that the ray velocity reciprocal has the same homogeneity degree as the slowness vector (i.e., zero). This proves that both, the ray velocity and its reciprocal, are zero-degree homogeneous functions wrt the tangent vector.

Since the numerator and the denominator of the proposed Lagrangian have the first and zero homogeneity degrees, respectively, the entire Lagrangian is a first-degree homogeneous function,

$$L(\mathbf{x}, k\mathbf{r}) = k L(\mathbf{x}, \mathbf{r}) \quad . \tag{13}$$

Furthermore, assume, for example, that the flow parameter is a scaled arclength, $s^* = s/k$. In this case,

$$\frac{d\mathbf{x}}{ds^*} = \frac{d\mathbf{x}}{ds}\frac{ds}{ds^*} = k\mathbf{r} \quad , \tag{14}$$

and,

$$\frac{\sqrt{d\mathbf{x}\cdot d\mathbf{x}}}{v_{\text{ray}}(\mathbf{x},\mathbf{r})} = \sqrt{\frac{d\mathbf{x}}{ds^*}\cdot\frac{d\mathbf{x}}{ds^*}}\frac{ds^*}{v_{\text{ray}}(\mathbf{x},k\mathbf{r})} = \frac{\sqrt{k\mathbf{r}\cdot k\mathbf{r}}}{v_{\text{ray}}(\mathbf{x},\mathbf{r})}\frac{ds}{k} = \frac{\sqrt{\mathbf{r}\cdot\mathbf{r}}\,ds}{v_{\text{ray}}(\mathbf{x},\mathbf{r})} = d\tau \quad , \tag{15}$$

i.e., the arclength scaling does not affect the traveltime. This is a general property of first-degree homogeneous parametric functionals: The choice of the flow parameter has no effect on the delivered value. In other words, the functional (in our case – the global traveltime $t$) is invariant under a change of parametric representation for the first-degree homogeneous integrand (Bliss, 1916, page 196). Throughout all parts of this study, we apply only the "natural" arclength of the path, $s$, and the normalized vector $\mathbf{r}$ tangent to the ray, $\mathbf{r}\cdot\mathbf{r} = 1$.

Alternative approach

An alternative approach has been suggested, for example, by Červený (2002b), where for the flow variable arclength, the Lagrangian is presented as (equation 68 of the cited work),

$$L(\mathbf{x},\mathbf{r}) = \frac{1}{v_{\text{ray}}(\mathbf{x},\mathbf{r})} \quad , \tag{16}$$

without the factor $\sqrt{\mathbf{r}\cdot\mathbf{r}}$ in the numerator. In this case, the "normalization" of the directional gradient of the ray velocity magnitude (required in our proposed method) is not needed; this gradient represents just a set of partial derivatives, $\partial v_{\text{ray}}/\partial \mathbf{r}$. The momentum equation holds in the following way,

$$L_{\mathbf{r}} = -\frac{1}{v_{\text{ray}}^2(\mathbf{x},\mathbf{r})}\frac{\partial v_{\text{ray}}}{\partial \mathbf{r}} = \mathbf{p} \quad . \tag{17}$$

Recall that for a general multivariable homogeneous function of degree $n$,

$$f(k\,\mathbf{r}) = k^n f(\mathbf{r}) \quad \text{(definition)} \quad , \tag{18}$$

where Euler's theorem holds (e.g., Buchanan and Yoon, 1999),

$$\nabla f(\mathbf{r})\cdot\mathbf{r} = n\, f(\mathbf{r}) \quad \text{(property)} \quad . \tag{19}$$

A converse statement is also true: A function with this property is homogeneous of degree $n$.

The property of the first-degree homogeneous function holds for the ray velocity reciprocal in the case of the non-normalized directional gradient of $\partial v_{\text{ray}}/\partial \mathbf{r}$,

$$\left(\frac{\partial}{\partial \mathbf{r}}\frac{1}{v_{\text{ray}}}\right)\cdot\mathbf{r} = -\frac{1}{v_{\text{ray}}^2(\mathbf{x},\mathbf{r})}\frac{\partial v_{\text{ray}}}{\partial \mathbf{r}}\cdot\mathbf{r} = \mathbf{p}\cdot\mathbf{r} = \frac{1}{v_{\text{ray}}} \quad , \tag{20}$$

and for the ray velocity magnitude itself,

$$\frac{\partial v_{\text{ray}}}{\partial \mathbf{r}} \cdot \mathbf{r} = -v_{\text{ray}}^2 \mathbf{p} \cdot \mathbf{r} = -v_{\text{ray}} \quad . \tag{21}$$

Equations 20 and 21 are particular cases of equation 19 for $n = +1$ and $n = -1$, respectively.

With this approach, both, the Lagrangian and the reciprocal of the ray velocity magnitude are first-order homogeneous functions wrt the ray velocity direction vector (unlike the case with our proposed Lagrangian where the homogeneity degree of the ray velocity is zero). The Lagrangian in equation 16 still leads to correct kinematic and dynamic equations. However, the disadvantage of such an approach is the (virtual) non-physical dependence of the ray velocity magnitude on the length of the tangent vector. Furthermore, in this case the ray velocity of isotropic media becomes also virtually dependent on the ray direction vector, with the same homogeneity degree as for a general anisotropic case, leading to $v_{\text{iso}} = v(\mathbf{x}, \mathbf{r})$ instead of $v_{\text{iso}} = v(\mathbf{x})$. Although this is a legitimate approach, we do not follow this way. We do apply the transform that converts the non-normalized directional gradient vector of the partial derivatives, $\partial v_{\text{ray}} / \partial \mathbf{r}$, into the normalized directional gradient, $\nabla_{\mathbf{r}} v_{\text{ray}} = \mathbf{T} \, \partial v_{\text{ray}} / \partial \mathbf{r}$, where $\mathbf{T} = \mathbf{I} - \mathbf{r} \otimes \mathbf{r}$, and in this case, Euler's theorem reduces to (equation A10 of Part I),

$$\nabla_{\mathbf{r}} v_{\text{ray}} \cdot \mathbf{r} = 0 \quad . \tag{22}$$

Equation 22 is a particular case of equation 19 for a zero-degree homogeneous function, $n = 0$.

Euler's theorem has been extended to higher derivatives (e.g., Shah and Sharma, 2014). In particular, for the second derivatives, equation 19 becomes,

$$\mathbf{r} \cdot \nabla \nabla f \cdot \mathbf{r} = n(n-1) f(\mathbf{r}) \quad . \tag{23}$$

Introducing the non-normalized directional gradient and Hessian of equations E2 and E6 of Part I, respectively, into equation 23, we obtain,

$$\mathbf{r} \cdot \frac{\partial^2 v_{\text{ray}}}{\partial \mathbf{r}^2} \cdot \mathbf{r} = \mathbf{r} \cdot \left( \frac{2}{v_{\text{ray}}} \frac{\partial v_{\text{ray}}}{\partial \mathbf{r}} \otimes \frac{\partial v_{\text{ray}}}{\partial \mathbf{r}} - v_{\text{ray}}^2 \frac{\partial \mathbf{p}}{\partial \mathbf{r}} \right) \cdot \mathbf{r}$$
$$= 2v_{\text{ray}}^3 \mathbf{r} \cdot (\mathbf{p} \otimes \mathbf{p}) \cdot \mathbf{r} - v_{\text{ray}}^2 \mathbf{r} \mathbf{p_r} \mathbf{r} = 2v_{\text{ray}}^3 (\mathbf{p} \cdot \mathbf{r})^2 - v_{\text{ray}}^2 \mathbf{r} \mathbf{p_r} \mathbf{r} \quad . \tag{24}$$

The second item on the right-hand side does not contribute as $\mathbf{p_r} \mathbf{r} = L_{\mathbf{rr}} \mathbf{r} = 0$ (the ray direction is the eigenvector of the positive semidefinite Lagrangian's Hessian, and the corresponding eigenvalue is zero). Equation 24 reduces to,

$$\mathbf{r} \cdot \frac{\partial^2 v_{\text{ray}}}{\partial \mathbf{r}^2} \cdot \mathbf{r} = 2 v_{\text{ray}} \quad , \tag{25}$$

as expected; for the homogeneity degree of the ray velocity, $n = -1$, equation 23 yields $n(n-1) = 2$. Now, consider the non-normalized Hessian of the ray velocity reciprocal, where the homogeneity degree of this reciprocal is $n = 1$. We obtain,

$$\mathbf{r} \cdot \left( \frac{\partial^2}{\partial \mathbf{r}^2} \frac{1}{v_{\text{ray}}} \right) \cdot \mathbf{r} = \mathbf{r} \cdot \left[ -\frac{\partial}{\partial \mathbf{r}} \left( \frac{1}{v_{\text{ray}}^2} \frac{\partial v_{\text{ray}}}{\partial \mathbf{r}} \right) \right] \cdot \mathbf{r} = \mathbf{r} \cdot \left( \frac{2}{v_{\text{ray}}^3} \frac{\partial v_{\text{ray}}}{\partial \mathbf{r}} \otimes \frac{\partial v_{\text{ray}}}{\partial \mathbf{r}} - \frac{1}{v_{\text{ray}}^2} \frac{\partial^2 v_{\text{ray}}}{\partial \mathbf{r}^2} \right) \cdot \mathbf{r} \quad . \tag{26}$$

The contribution of the second item on the right-hand side is known from equation 26, and this leads to,

$$\mathbf{r} \cdot \left( \frac{\partial^2}{\partial \mathbf{r}^2} \frac{1}{v_{\text{ray}}} \right) \cdot \mathbf{r} = \frac{2}{v_{\text{ray}}^3} \mathbf{r} \cdot \left( \frac{\partial v_{\text{ray}}}{\partial \mathbf{r}} \otimes \frac{\partial v_{\text{ray}}}{\partial \mathbf{r}} \right) \cdot \mathbf{r} - \frac{2}{v_{\text{ray}}}$$
$$= 2 v_{\text{ray}} \mathbf{r} \cdot (\mathbf{p} \otimes \mathbf{p}) \cdot \mathbf{r} - \frac{2}{v_{\text{ray}}} = 2 v_{\text{ray}} (\mathbf{p} \cdot \mathbf{r})^2 - \frac{2}{v_{\text{ray}}} = 0 \quad , \tag{27}$$

as expected; for $n=1$, equation 23 yields $n(n-1)=0$.

For the normalized Hessian of the ray velocity, the homogeneity degree of the velocity is zero, $n=0$, and $n(n-1)=0$. Applying equation A12 of Part I, we obtain the expected result,

$$\mathbf{r} \cdot \nabla_{\mathbf{r}} \nabla_{\mathbf{r}} v_{\text{ray}} \cdot \mathbf{r} = 0 \qquad . \tag{28}$$

Concluding remark

Taking into account equations 21 and 22, we conclude that the ray velocity magnitude $v_{\text{ray}}(\mathbf{x}, \mathbf{r})$ can be considered as a homogeneous function of either degree zero or degree minus one wrt the ray direction vector, $\mathbf{r}$, depending on the definition of the directional gradient of the ray velocity magnitude, $v_{\text{ray}}$. If the directional gradient of the ray velocity magnitude is the "non-normalized" vector $\partial v_{\text{ray}} / \partial \mathbf{r}$ (in this case, the Lagrangian has only unity in the numerator), then $v_{\text{ray}}(\mathbf{x}, \mathbf{r})$ is a homogeneous function of degree minus one. Otherwise, if the directional gradient is the "normalized" vector $\nabla_{\mathbf{r}} v_{\text{ray}}$ (in this case, the Lagrangian should include $\sqrt{\mathbf{r} \cdot \mathbf{r}}$ in the numerator), then $v_{\text{ray}}(\mathbf{x}, \mathbf{r})$ is a zero-degree homogeneous function. There is no contradiction, just two different approaches. However, we prefer our approach, since unlike the common form of the alternative Lagrangian, the proposed Lagrangian strictly obeys the fundamental physical characteristics of wave propagation in isotropic and anisotropic elastic continua:

a) The anisotropic ray velocity magnitude is independent of the length of the tangent vector, $k$; it depends only on the direction $\mathbf{r}$ of this vector and the medium properties at the given position $\mathbf{x}$.

b) The isotropic velocity becomes naturally only a function of the ray position, $v_{iso} = v(\mathbf{x})$, and hence, applying operator $\nabla_\mathbf{r}$ to establish any directional derivatives in isotropic media leads to a vanishing result.

Both items (a and b) do not hold in the abovementioned alternative approach.

We finally note that in both cases (the Eigenray method and the alternative approach), the Lagrangian is the first-degree homogeneous function; Červený (2002a, 2002b) suggested also a modified, second-degree homogeneous Lagrangian, and we discuss it later in this part of the study.

## OTHER ALTERNATIVE LAGRANGIANS

Different forms of Lagrangians can be used to obtain the required stationary solution. The form of the Lagrangian depends on the flow parameter along the ray, which can be, for example, traveltime $d\tau$, arclength $ds = v_{ray}(\tau) d\tau$ (as proposed in our Eigenray method), or parameter sigma $d\sigma = v_{ray}^2(\tau) d\tau$, where $v_{ray}$ is the magnitude of the ray velocity at a given value of the flow parameter. Note that even for the same flow parameter, multiple co-existing forms of Lagrangian are possible.

<u>Alternative forms of the Lagrangian and the Finsler metric tensor</u>

Červený (2002a, 2002b) considers several forms of the Lagrangian for a general flow parameter. In particular, for the flow parameter traveltime, Červený suggests the following two Lagrangians,

$$L^U\left(\mathbf{x},\dot{\mathbf{x}}_\tau\right)=\sqrt{\dot{\mathbf{x}}_\tau\,\mathbf{G}\,\dot{\mathbf{x}}_\tau} \quad \text{and} \quad L^M\left(\mathbf{x},\dot{\mathbf{x}}_\tau\right)=\frac{1}{2}\dot{\mathbf{x}}_\tau\,\mathbf{G}\,\dot{\mathbf{x}}_\tau \quad ; \qquad \dot{\mathbf{x}}_\tau=\mathbf{v}_{\text{ray}} \quad , \quad (29)$$

where $\mathbf{G}(\mathbf{x},\mathbf{v}_{\text{ray}})$ is the Finsler metric tensor. The unmodified Lagrangian, $L^U$, is of the first-degree homogeneity wrt the ray velocity $\dot{\mathbf{x}}_\tau$, while the modified Lagrangian, $L^M$, is of the second degree homogeneity.

For the modified Lagrangian, $L^M$, Červený relates the inverse of the Finsler metric $\mathbf{G}^{-1}$ to the Hessian of the corresponding eigenvalue Hamiltonian $H^\lambda$ wrt the slowness vector, $H^\lambda_{\mathbf{pp}}$. This Hamiltonian, in turn, is defined as the unit eigenvalue $\lambda$ of the Christoffel matrix (tensor) (Appendix A); therefore, we name it "the eigenvalue Hamiltonian", to distinguish from the other Hamiltonians.

In addition, the Finsler metric $\mathbf{G}$ directly relates the slowness vector to the ray velocity vector. However, this relationship is not unique. Following, for example, Musgrave (1970), Červený (1972), and Tsvankin (2012), a different second-order tensor $\mathbf{F}$, $\mathbf{F}\neq\mathbf{G}^{-1}$, relating the slowness and the ray velocity vectors, can be used, which is a function of the density-normalized fourth-order stiffness tensor $\tilde{\mathbf{C}}$ and the polarization vector $\mathbf{g}$. We discuss these relations in Appendix C. However, we note that using the Lagrangians $L^U$ and $L^M$ imposes computing the first and second derivatives of the polarization vector $\mathbf{g}$ wrt the slowness components, or the second derivatives of the Christoffel eigenvalue $\lambda$, which is an attainable but not a straightforward procedure. This is one of the main reason for choosing our proposed Lagrangian, which does not explicitly depend on the polarization. We therefore consider our proposed Lagrangian more convenient and efficient than the other known alternatives.

## HAMILTONIAN RELATED TO THE MODIFIED LAGRANGIAN

Červený (2002a, 2002b) relates the eigenvalue Hamiltonian $H^\lambda(\mathbf{x},\mathbf{p})$ to the modified Lagrangian $L^M(\mathbf{x},\dot{\mathbf{x}}_\tau)$. Unlike the vanishing reference Hamiltonian, defined in equation 11 of Part I and representing the Christoffel equation, the eigenvalue Hamiltonian is the root of this equation, i.e., the unit eigenvalue, $\lambda = 1$, of the Christoffel matrix (up to the factor $1/2$),

$$H^\lambda(\mathbf{x},\mathbf{p}) \equiv H^\lambda\left[\mathbf{x},\mathbf{p},\mathbf{g}(\mathbf{x},\mathbf{p})\right] = \frac{1}{2}\mathbf{g}\left(\mathbf{p}\tilde{\mathbf{C}}\mathbf{p}\right)\mathbf{g} = \frac{1}{2}\mathbf{g}\mathbf{\Gamma}\mathbf{g} = \frac{\lambda}{2} = \frac{1}{2} \quad , \qquad (30)$$

where, $\mathbf{g}\mathbf{\Gamma}\mathbf{g} = \lambda\mathbf{g}\cdot\mathbf{g} = \lambda$, as the polarization vector $\mathbf{g}$ is normalized to the unit length. In this Hamiltonian equation, $\mathbf{\Gamma}(\mathbf{x},\mathbf{p}) = \mathbf{p}\tilde{\mathbf{C}}(\mathbf{x})\mathbf{p}$ is the Christoffel matrix (tensor), $\mathbf{g}(\mathbf{x},\mathbf{p})$ is the polarization vector (the eigenvector of the Christoffel matrix), and $\tilde{\mathbf{C}}(\mathbf{x})$ is the density-normalized fourth-order stiffness tensor. The tilde recalls that it is the stiffness tensor of order four, rather than the Voigt or Kelvin matrix representation. The eigenvalue Hamiltonian depends on the polarization vector, $\mathbf{g}$. It can be computed, given the medium properties at the location $\mathbf{x}$ and the slowness vector $\mathbf{p}$ or its direction $\mathbf{n}$ (the phase direction).

## LEGENDRE TRANSFORM

Equation set 15 of Part I represents the conventional Hamiltonian kinematic ray tracing relationships. In particular, similar kinematic relationships are given by Červený (2002b), in equation 29 of the cited paper, where the flow parameter is the traveltime $\tau$, and the eigenvalue Hamiltonian $H^\lambda(\mathbf{x},\mathbf{p})$ defined in equation 30 is used. This eigenvalue Hamiltonian matches the Lagrangian $L^M(\mathbf{x}, \dot{\mathbf{x}}_\tau)$ that was modified to be a homogeneous function of the second degree wrt $\dot{\mathbf{x}}_\tau \equiv d\mathbf{x}/d\tau$, whereas in our case the Lagrangian is a homogeneous function of the first degree wrt $\mathbf{r} \equiv \dot{\mathbf{x}} \equiv d\mathbf{x}/ds$. We follow equation 42 of the cited paper, where the modified Lagrangian, $L^M[\mathbf{x}(\tau), \dot{\mathbf{x}}_\tau(\tau)]$, defined in equation 29, is related to its corresponding Hamiltonian, $H^\lambda[\mathbf{x}(\tau), \mathbf{p}(\tau)]$, defined in equation 30, through the well-known Legendre transform (e.g., Arnold, 1989; Slawinski, 2015; Červený, 2002a; Červený 2002b). We emphasize that flow variable of $L^M$ and $H^\lambda$ is the traveltime, $\tau$.

In a similar way, the proposed Lagrangian, $L[\mathbf{x}(s), \mathbf{r}(s)]$, defined in equation 1, can be connected to its corresponding arclength-related Hamiltonian, $H[\mathbf{x}(s), \mathbf{p}(s)]$, defined in equations 11 and 13 of Part I,

$$\underbrace{L^M(\mathbf{x}, \dot{\mathbf{x}}_\tau)}_{1/2} = \underbrace{\dot{\mathbf{x}}_\tau \cdot \mathbf{p}}_{1} - \underbrace{H^\lambda(\mathbf{x}, \mathbf{p})}_{1/2} \quad , \quad \text{where} \quad \underbrace{L^M_{\dot{\mathbf{x}}_\tau} = \mathbf{p}}_{\text{momentum}} \quad \text{and} \quad L^M_{\mathbf{x}} = \frac{d\mathbf{p}}{d\tau} \quad ,$$

$$\underbrace{L(\mathbf{x}, \mathbf{r})}_{v_{\text{ray}}^{-1}} = \underbrace{\mathbf{r} \cdot \mathbf{p}}_{v_{\text{ray}}^{-1}} - \underbrace{H(\mathbf{x}, \mathbf{p})}_{\text{zero}} \quad , \quad \text{where} \quad \underbrace{L_\mathbf{r} = \mathbf{p}}_{\text{momentum}} \quad \text{and} \quad L_\mathbf{x} = \frac{d\mathbf{p}}{ds} \quad . \tag{31}$$

Recall that the modified Lagrangian $L^M(\mathbf{x}, \dot{\mathbf{x}}_\tau)$ is a second-degree homogeneous function wrt the ray velocity vector $\dot{\mathbf{x}}_\tau = \mathbf{v}_{\text{ray}}$, while the proposed arclength-related Lagrangian is a first-degree homogenous function wrt the ray velocity direction $\mathbf{r}$. Hence, it follows from Euler's theorem,

$$\dot{\mathbf{x}}_\tau \cdot \mathbf{p} = \dot{\mathbf{x}}_\tau \cdot L^M_{\dot{\mathbf{x}}_\tau} = 2L^M(\mathbf{x}, \dot{\mathbf{x}}_\tau) \quad , \quad \mathbf{r} \cdot \mathbf{p} = \mathbf{r} \cdot L_\mathbf{r} = L(\mathbf{x}, \dot{\mathbf{x}}_\tau) \quad . \tag{32}$$

Combining equations 31 and 32, we obtain,

$$L^M(\mathbf{x},\dot{\mathbf{x}}_\tau) = H^\lambda(\mathbf{x},\mathbf{p}) \quad \text{and} \quad H(\mathbf{x},\mathbf{p}) = 0 \quad , \tag{33}$$

which is in agreement with both, the eigenvalue and arclength-related Hamiltonians.

Since the arclength-related Hamiltonian $H(s)$ vanishes along the ray, the value of the corresponding Lagrangian $L(s)$ is,

$$L(s) = \mathbf{r}(s) \cdot \mathbf{p}(s) = \frac{1}{v_{\text{ray}}(s)} \quad , \tag{34}$$

and the factor $\sqrt{\mathbf{r} \cdot \mathbf{r}}$ should be set in the numerator (on the right-hand side of equation 34) in the case where the directional gradient, $L_{\mathbf{r}}$, or Hessian, $L_{\mathbf{rr}}$, (or the mixed Hessians, $L_{\mathbf{xr}} = L_{\mathbf{rx}}^T$) of the Lagrangian have to be computed.

In Figure 1, we present the well-known relations between the Hamiltonian and Lagrangian approaches for both ray theory and particle mechanics, where $E_p$ and $E_k$ are the potential and kinetic energy, respectively (e.g., Arnold, 1989; Slawinski, 2015). We summarized the notations and definitions for the Hamiltonians and Lagrangians used in all parts of this study in Tables 1 and 2 of Part I. Both, the Lagrangian and its matching Hamiltonian, have the units of time, divided by the units of the flow variable, $\left[H^\varsigma\right] = \left[L^\varsigma\right] = \left[T/\varsigma\right]$. Accordingly, for the flow variable traveltime, they are unitless, and for the arclength they have the units of slowness.

## THE GENERALIZED MOMENTUM EQUATION

The generalized momentum equation, $L_\mathbf{r} = \mathbf{p}$, derived in Part I, can be (in principle) used to find the ray direction components vs. the slowness vector, $\mathbf{r} = \mathbf{r}(\mathbf{x}, \mathbf{p})$. We emphasize that this is not the optimum way to establish the ray direction, as the direct explicit Hamiltonian solution exists, $\mathbf{r} = H_\mathbf{p}(\mathbf{x}, \mathbf{p})$ (this is one of the kinematic equations, where the ray direction vector is, $\mathbf{r} = \dot{\mathbf{x}} = d\mathbf{x}/ds$). For completeness, in Appendix H we show that this problem (finding $\mathbf{p}$ for a given $\mathbf{r}$) can be also solved (numerically) with the momentum equation, $L_\mathbf{r}(\mathbf{x}, \mathbf{r}) = \mathbf{p}$, using the proposed Lagrangian $L$ and an additional physical constraint enforcing the normalization of the ray velocity direction vector, $\mathbf{r} \cdot \mathbf{r} = 1$. The constraint is needed to remove the singularity of the directional Hessian matrix $L_{\mathbf{rr}}(\mathbf{x}, \mathbf{r})$ of the proposed Lagrangian, which is only first-degree homogeneous wrt the ray direction $\mathbf{r}$. Furthermore, the momentum equation, $L^U_{\dot{\mathbf{x}}_\tau}(\mathbf{x}, \dot{\mathbf{x}}_\tau) = \mathbf{p}$, based on the unmodified first-order homogeneity Lagrangian $L^U(\tau)$ suggested by Červený (2002a, 2002b) can be also used, in this case to find the ray velocity vector $\dot{\mathbf{x}}_\tau = \mathbf{v}_{\text{ray}}$, where the scalar normalization condition is, $\dot{\mathbf{x}}_\tau \cdot \mathbf{p} = 1$.

The inverse problem of finding the slowness vector $\mathbf{p}$, given the ray direction vector $\mathbf{r}$, is solved in Part I. Recall that the governing resolving conditions are: a) the slowness gradient of the reference Hamiltonian, $H^{\bar{\tau}}_\mathbf{p}$, is collinear with the ray direction, and b) the Hamiltonian accepts a known constant value; in our case it vanishes, $H^{\bar{\tau}}(\mathbf{x}, \mathbf{p}) = 0$. Any Hamiltonian can be applied; the reference Hamiltonian is the simplest for this specific problem.

**VALIDATION OF THE PROPOSED ARCLENGTH-RELATED LAGRANGIAN**

In order to validate the correctness of the proposed Lagrangian $L(\mathbf{x},\mathbf{r})$ for general anisotropy, we first prove analytically in Appendix D that the proposed Lagrangian, $L(\mathbf{x},\mathbf{r})$, and the two Lagrangians, $L^U(\mathbf{x},\dot{\mathbf{x}}_\tau)$ and $L^M(\mathbf{x},\dot{\mathbf{x}}_\tau)$, unmodified and modified, respectively, suggested by Červený (2002a, 2002b), lead to identical kinematic ray equations. We then demonstrate equivalent results obtained with three different anisotropic examples (two numerical and one analytical, Appendices E, F, G), where the arclength derivative of the ray position vector, $\mathbf{r}=d\mathbf{x}/ds=\dot{\mathbf{x}}$, and that of the slowness vector, $d\mathbf{p}/ds=\dot{\mathbf{p}}$, are computed in three different ways: a) with the proposed, first-degree homogeneity in $\mathbf{r}$, Lagrangian $L(s)$, b) with its corresponding arclength-related Hamiltonian $H(s)$, connected to $L(s)$ through the Legendre transform (equation 31), and c) with the eigenvalue Hamiltonian $H^\lambda(\tau)$ connected to the modified, second-degree homogeneity in $\dot{\mathbf{x}}_\tau$, Lagrangian $L^M(\tau)$ (equation 29). In Appendix E we use a tilted orthorhombic medium, and in Appendix F we use the most general anisotropic case of a spatially varying triclinic medium with all 21 stiffness components. In Appendix G we study analytically an ellipsoidal anisotropic medium. We demonstrate that the results obtained by the three methods are equal, which means that they lead to identical kinematic ray tracing equations.

## CONCLUSIONS

In this part of the study, we validate our proposed Lagrangian and compare it with alternative Lagrangians. The main advantage of the proposed Lagrangian is the clear physical meaning of the directional dependencies of its associated ray velocities for anisotropic elastic media, which naturally reduces to direction-independent velocities in isotropic case. We prove that the kinematic equations that follow from our arclength-related Lagrangian and from the time-related alternative Lagrangians are identical. We provide the Legendre transform connecting the proposed arclength-related Lagrangian to its corresponding Hamiltonian. The proposed Lagrangian is validated analytically for a canonical case of ellipsoidal anisotropic medium, and numerically for spatially varying orthorhombic and general triclinic media. The validation process involves computation of the derivative of the slowness vector wrt the arclength, applying the Lagrangian approach and two different Hamiltonian approaches. Finally, we demonstrate that, unlike the common consideration, a second-degree homogeneous Lagrangian with respect to the vector tangent to the ray (where its directional Hessian matrix is non-singular), is not a must in the Lagrangian-based ray bending method.

**ACKNOWLEDGEMENT**

The authors are grateful to Emerson for the financial and technical support of this study and for the permission to publish its results. The gratitude is extended to Ivan Pšenčík, Einar Iversen, Michael Slawinski, Alexey Stovas, Vladimir Grechka, and our colleague Beth Orshalimy, whose valuable remarks helped to improve the content and style of this paper.

**APPENDIX A. MODIFIED SECOND-DEGREE HEOMOGENEOUS LAGRANGIAN AND ITS MATCHING EIGENVALUE HAMILTONIAN**

Červený (2002b), in equations 20 and 41 of the cited paper, defines the stationary traveltime integral in the two following forms,

$$t = \int_S^R L^U(\tau) d\tau = \int_S^R \sqrt{\dot{\mathbf{x}}_\tau \mathbf{G} \dot{\mathbf{x}}_\tau}\, d\tau \;,$$
$$t = 2\int_S^R L^M(\tau) d\tau = \int_S^R \dot{\mathbf{x}}_\tau \mathbf{G} \dot{\mathbf{x}}_\tau\, d\tau \;,$$

$$\mathbf{v}_{\text{ray}} \equiv \dot{\mathbf{x}}_\tau \equiv \frac{d\mathbf{x}}{d\tau} \;,$$
$$\text{stationarity}: \delta t \to 0 \;,$$
(A1)

where $L^U(\mathbf{x}, \dot{\mathbf{x}}_\tau)$ and $L^M(\mathbf{x}, \dot{\mathbf{x}}_\tau)$ are the unmodified and modified Lagrangians, of the first- and second-degree homogeneity, respectively, wrt the ray velocity components, $v_{\text{ray},i} = \dot{x}_{\tau,i}, i = 1, 2, 3$. Tensor $\mathbf{G}(\mathbf{x}, \dot{\mathbf{x}}_\tau)$ is the Finsler metric (also called the propagation metric tensor), accounting for the dependency of the ray velocity on the position and direction along the ray. It has the units of slowness squared, $\left[\text{T}^2 / \text{L}^2\right]$. According to the two cited papers, the Finsler metric tensor $\mathbf{G}$, can be defined through its inverse (e.g., Červený, 2002b, equation 53),

$$\mathbf{G}^{-1} = H^\lambda_{\mathbf{pp}}, \quad \mathbf{G} = \left(H^\lambda_{\mathbf{pp}}\right)^{-1} \tag{A2}$$

where $H^\lambda_{\mathbf{pp}}$ is the second derivative of the eigenvalue Hamiltonian, $H^\lambda(\mathbf{x}, \mathbf{p})$ (defined in equation 30 and listed in Table 1 of Part I) wrt the slowness vector components. We use here the same shorthand notation for the Hessian as that defined in equation 9 of Part I for the gradients,

$$H^\lambda_{\mathbf{pp}}(\mathbf{x}, \mathbf{p}) = \nabla_{\mathbf{p}} \nabla_{\mathbf{p}} H^\lambda(\mathbf{x}, \mathbf{p}) = \frac{\partial^2}{\partial \mathbf{p}^2} H^\lambda(\mathbf{x}, \mathbf{p}) \quad . \tag{A3}$$

Thus, the ability to use the Lagrangian of equation A1 in general anisotropic (e.g., triclinic) elastic media depends on the feasibility to compute the Finsler metric tensor $\mathbf{G}$, representing the inverse of $H^\lambda_{\mathbf{pp}}$, which, in turn, is a function of the polarization vector $\mathbf{g}$; note that $\mathbf{g} = \mathbf{g}(\mathbf{x}, \mathbf{p})$ is an implicit function of the location and slowness components. Hence, establishing $H_{\mathbf{pp}}^{\lambda}$ may require the computation of the first and second partial derivatives of the polarization vector **g** or those derivatives of the eigenvalue $\lambda$ wrt the slowness components, whose analytical computation is challenging. We are unaware of any suitable relationship for obtaining the matrix (tensor) $\partial \mathbf{g} / \partial \mathbf{p}$ and the third-order tensor $\partial^2 \mathbf{g} / \partial \mathbf{p}^2$. Moreover, the numerical computation of these partial derivatives is questionable as we cannot perturb one slowness component without a simultaneous change in the two other components (this leads to a violation of the Christoffel equation). In other words, we cannot compute the polarization at the same location, for a neighbor slowness direction, where one slowness component obtains a small finite increment, and the two others remain fixed.

## APPENDIX B.
## RAY TRACING EQUATIONS WITH THE EIGENVALUE HAMILTONIAN

As mentioned, Červený (2002a, 2002b) suggests using the eigenvalue Hamiltonian, $H^{\lambda}\left[\mathbf{x}(\tau), \mathbf{p}(\tau)\right]$, defined in equation 30 and corresponding to the modified, second-degree homogeneous Lagrangian $L^{M}\left[\mathbf{x}(\tau), \mathbf{v}_{\text{ray}}(\tau)\right]$ of equation 29. The flow variable of the eigenvalue Hamiltonian is traveltime, and the kinematic equations read,

$$\frac{d\mathbf{x}}{d\tau} = \frac{\partial H^{\lambda}}{\partial \mathbf{p}} \quad , \quad \frac{d\mathbf{p}}{d\tau} = -\frac{\partial H^{\lambda}}{\partial \mathbf{x}} \quad , \tag{B1}$$

or equivalently,

$$\frac{d\mathbf{x}}{d\tau} = \frac{1}{2}\frac{\partial \lambda}{\partial \mathbf{p}} \quad , \qquad \frac{d\mathbf{p}}{d\tau} = -\frac{1}{2}\frac{\partial \lambda}{\partial \mathbf{x}} \quad , \tag{B2}$$

where $\lambda = 1$ is the unit eigenvalue of the Christoffel matrix (for compressional waves). Then, according to Červený (2000), equation 3.6.8,

$$\frac{\partial \lambda}{\partial p_i} = \frac{\partial \Gamma_{jl}}{\partial p_i} g_j g_l \quad , \qquad \frac{\partial \lambda}{\partial x_i} = \frac{\partial \Gamma_{jl}}{\partial x_i} g_j g_l \quad , \tag{B3}$$

and,

$$\frac{dx_i}{d\tau} = \frac{1}{2}\frac{\partial \Gamma_{jl}}{\partial p_i} g_j g_l \quad , \qquad \frac{dp_i}{d\tau} = -\frac{1}{2}\frac{\partial \Gamma_{jl}}{\partial x_i} g_j g_l \quad , \tag{B4}$$

where (Červený, 2000, equation 3.6.9),

$$\frac{\partial \Gamma_{jl}}{\partial p_i} = \left(\tilde{C}_{ijkl} + \tilde{C}_{jkli}\right) p_k \quad , \qquad \frac{\partial \Gamma_{jl}}{\partial x_i} = \frac{\partial \tilde{C}_{jkln}}{\partial x_i} p_k p_n \quad . \tag{B5}$$

In tensor notations we can write,

$$\frac{\partial \mathbf{\Gamma}}{\partial \mathbf{p}} = \frac{\partial}{\partial \mathbf{p}}\left(\mathbf{p}\tilde{\mathbf{C}}\mathbf{p}\right) = \mathbf{p}\tilde{\mathbf{C}} + \tilde{\mathbf{C}}\mathbf{p} \quad . \tag{B6}$$

Note that $\partial \mathbf{\Gamma}/\partial \mathbf{p}$ is a third-order tensor, thus, each component is defined by three indices. Two indices belong to the Christoffel tensor $\mathbf{\Gamma}$, and one index points to the Cartesian component of the gradient. In this case, indices 1 and 3 belong to $\mathbf{\Gamma}$, and index 2 belongs to the gradient components. The ray velocity vector reads,

$$\mathbf{v}_{\text{ray}} = \frac{d\mathbf{x}}{d\tau} = \frac{1}{2}\mathbf{g}\left(\mathbf{p}\tilde{\mathbf{C}} + \tilde{\mathbf{C}}\mathbf{p}\right)\mathbf{g} \quad . \tag{B7}$$

Due to the symmetry of the stiffness tensor, the two items in the sum in the equation above are equal,

$$\mathbf{g}(\mathbf{p}\tilde{\mathbf{C}})\mathbf{g} = \mathbf{g}(\tilde{\mathbf{C}}\mathbf{p})\mathbf{g} = (\mathbf{g}\tilde{\mathbf{C}}\mathbf{g})\mathbf{p} \qquad , \tag{B8}$$

so that,

$$\mathbf{v}_{\text{ray}} = \dot{\mathbf{x}}_\tau = \frac{d\mathbf{x}}{d\tau} = \frac{1}{2}\mathbf{g}(\mathbf{p}\tilde{\mathbf{C}} + \tilde{\mathbf{C}}\mathbf{p})\mathbf{g} = \mathbf{g}\tilde{\mathbf{C}}\mathbf{g}\mathbf{p} \qquad . \tag{B9}$$

Thus, the first kinematic ray tracing equation reads,

$$\mathbf{r} = \frac{d\mathbf{x}}{ds} = \frac{\mathbf{g}\tilde{\mathbf{C}}\mathbf{g}\mathbf{p}}{v_{\text{ray}}} \qquad \text{or} \qquad \mathbf{r} = \frac{\mathbf{g}(\mathbf{p}\tilde{\mathbf{C}} + \tilde{\mathbf{C}}\mathbf{p})\mathbf{g}}{2v_{\text{ray}}} \qquad . \tag{B10}$$

The second kinematic ray tracing equation reads (e.g., Červený, 2000, equations 3.6.12 and 3.6.15),

$$\frac{d\mathbf{p}}{d\tau} = -\frac{1}{2}\frac{\partial \tilde{C}_{jkln}}{\partial x_i} p_k p_n g_j g_l \qquad . \tag{B11}$$

Tensor $\nabla_{\mathbf{x}}\tilde{C}_{jklni} = \dfrac{\partial \tilde{C}_{jkln}}{\partial x_i}$ is of order five, and its last index is the index of the gradient component. For the sake of symmetry, we move the index of the gradient component to the central position (the third from the total five). In Wolfram Mathematica, this transpose operator looks like,

$$\nabla_{\mathbf{x}}^T \tilde{\mathbf{C}} = \text{Transpose}\left[\nabla_{\mathbf{x}}\tilde{\mathbf{C}}, \{1,2,4,5,3\}\right] \qquad . \tag{B12}$$

That is to say, indices 3 and 4 become 4 and 5, respectively, and index 5 becomes 3. Then, the second ray tracing equation can be arranged as,

$$\frac{d\mathbf{p}}{d\tau} = -\frac{1}{2}\mathbf{p}\left(\mathbf{g}\nabla_\mathbf{x}^T \tilde{\mathbf{C}}\mathbf{g}\right)\mathbf{p} \quad , \tag{B13}$$

or for the flow variable arclength,

$$\frac{d\mathbf{p}}{ds} = -\frac{\mathbf{p}\left(\mathbf{g}\nabla_\mathbf{x}^T \tilde{\mathbf{C}}\mathbf{g}\right)\mathbf{p}}{2v_{\text{ray}}} \quad . \tag{B14}$$

Due to the symmetry of the stiffness tensor (and its spatial gradient), this can be also arranged as,

$$\frac{d\mathbf{p}}{ds} = -\frac{\mathbf{g}\left(\mathbf{p}\nabla_\mathbf{x}^T \tilde{\mathbf{C}}\mathbf{p}\right)\mathbf{g}}{2v_{\text{ray}}} \quad . \tag{B15}$$

This relationship makes it possible to obtain the arclength derivative of the slowness vector, $d\mathbf{p}/ds$ from the eigenvalue Hamiltonian defined in equation 30. Equation B15 will be later used to validate the correctness of our proposed Lagrangian that can be also applied to compute $d\mathbf{p}/ds$.

### APPENDIX C. SLOWNESS VS. RAY VELOCITY RELATIONSHIPS

The Finsler metric tensor $\mathbf{G}$ can be defined in both, slowness (phase) and ray velocity domains, as $\mathbf{G}(\mathbf{x},\mathbf{p})$ and $\mathbf{G}(\mathbf{x},\dot{\mathbf{x}}_\tau)$ and makes it possible to relate the slowness vector $\mathbf{p}$ to the ray velocity vector $\mathbf{v}_{\text{ray}}$ in general anisotropic media (Červený, 2002b, equations 58 and 59),

$$\mathbf{p} = \mathbf{G}\dot{\mathbf{x}}_\tau \quad , \quad \dot{\mathbf{x}}_\tau = \mathbf{G}^{-1}\mathbf{p} \quad ,$$
$$\mathbf{G}^{-1} = H^\lambda_{\mathbf{pp}} \quad , \quad \dot{\mathbf{x}}_\tau = \mathbf{v}_{\text{ray}} \quad .$$
(C1)

Alternatively, the ray velocity and the slowness vectors can be related through the stiffness tensor $\tilde{\mathbf{C}}$ and the polarization vector $\mathbf{g}$ (e.g., Fedorov, 1968; Musgrave, 1970; Červený, 1972; Auld, 1973; Tsvankin, 2012),

$$\mathbf{v}_{\text{ray}} = \mathbf{F}\mathbf{p} \quad , \quad \text{where} \quad \mathbf{F} = \mathbf{g}\tilde{\mathbf{C}}\mathbf{g} \quad .$$
(C2)

To establish the conversion matrix $\mathbf{F}$, we need to solve the Christoffel equation 11 of Part I for the phase velocity, $v_{\text{phs}}$, related to the specified wave type, given the slowness direction $\mathbf{n}$, and to find the polarization vector $\mathbf{g}$ using the Christoffel matrix,

$$\hat{\mathbf{\Gamma}} = \mathbf{n}\tilde{\mathbf{C}}\mathbf{n} \quad , \quad \det\left(\hat{\mathbf{\Gamma}} - v^2_{\text{phs}}\right) = 0 \quad , \quad \hat{\mathbf{\Gamma}}\mathbf{g} = v^2_{\text{phs}}\mathbf{g} \quad .$$
(C3)

The inverse relationship from the ray velocity vector to the slowness vector also exists, provided matrix $\mathbf{F} = \mathbf{g}\tilde{\mathbf{C}}\mathbf{g}$ is invertible, i.e., $\det \mathbf{F} \neq 0$,

$$\mathbf{p} = \mathbf{F}^{-1}\mathbf{v}_{\text{ray}} \quad .$$
(C4)

Note that although equation C4 represents a valid relationship, this is normally not the way to find the slowness vector. First, in the Eigenray workflow, we only know the direction $\mathbf{r}$ of the ray velocity, while its magnitude $v_{\text{ray}}$ is unknown and should be computed. To compute the ray velocity magnitude, we need first to establish the slowness vector $\mathbf{p}$ (which makes equation C4 unnecessary). Second, to compute matrix $\mathbf{F}$, we need the polarization vector $\mathbf{g}$ which is unknown either. Recall that given the ray direction $\mathbf{r}$, the slowness $\mathbf{p}$ is defined with equation set 18 of Part I that exploits the reference Hamiltonian $H^{\bar{\tau}}(\mathbf{x},\mathbf{p})$ and its slowness gradient $H^{\bar{\tau}}_{\mathbf{p}}$. Then the ray velocity magnitude can be established with equation 34, and the polarization (if needed) – with the Christoffel matrix $\Gamma(\mathbf{x},\mathbf{p})$.

We emphasize (and demonstrate in Appendix G) that although both equations C1 and C2 are valid, the inverse Finsler metric $\mathbf{G}^{-1}$ and tensor $\mathbf{F}$ are different tensors. Subtracting equation C2 from the upper equation in the right column of set C1, we obtain,

$$\begin{aligned}\mathbf{v}_{\text{ray}} &= \mathbf{G}^{-1}\mathbf{p} \\ \mathbf{v}_{\text{ray}} &= (\mathbf{g}\tilde{\mathbf{C}}\mathbf{g})\mathbf{p}\end{aligned} \quad , \quad (\mathbf{G}^{-1} - \mathbf{g}\tilde{\mathbf{C}}\mathbf{g})\mathbf{p} = 0 \quad . \tag{C5}$$

The difference in the brackets represents a matrix whose three rows are coplanar vectors and belong to a plane normal to the slowness vector (and hence they are dependent). It follows from equation C5 that the slowness direction is an eigenvector of the matrix $\mathbf{A} \equiv \mathbf{G}^{-1} - \mathbf{g}\tilde{\mathbf{C}}\mathbf{g}$, where $\det \mathbf{A} = 0$, and the corresponding eigenvalue for this eigenvector is zero.

The matrix (tensor) $\mathbf{F} = \mathbf{g}\tilde{\mathbf{C}}\mathbf{g}$ always exists, and it is positive definite provided the Christoffel matrix (tensor) $\Gamma = \mathbf{p}\tilde{\mathbf{C}}\mathbf{p}$ is also positive definite, which is the case for any anisotropy except the acoustic approximation, where both $\Gamma$ and $\mathbf{F}$ become positive semidefinite, with vanishing determinants, due to the vanishing shear velocities in one or more directions. In Appendix I we consider a special case of the matrix $\mathbf{F}$ with a vanishing determinant. However, even in this case, the Finsler metric $\mathbf{G} = \left(H^{\lambda}_{\mathbf{pp}}\right)^{-1}$ may exist (we recall that the eigenvalue Hamiltonian $H^{\lambda}$ is defined in equation 30).

As mentioned above, the Christoffel matrix can be also presented as $\hat{\mathbf{\Gamma}} = \mathbf{n}\tilde{\mathbf{C}}\mathbf{n}$, where $\mathbf{n}$ is the slowness direction, $\mathbf{n} = \mathbf{p}/\sqrt{\mathbf{p}\cdot\mathbf{p}}$. The corresponding eigenvalues are the phase velocities squared,

$$\lambda = \begin{bmatrix} v_{\text{phs},P}^2 & v_{\text{phs},S1}^2 & v_{\text{phs},S2}^2 \end{bmatrix} \quad . \tag{C6}$$

The subscripts P, S1 and S2 indicate compressional, fast shear and slow shear waves, respectively. The eigenvalues of the Christoffel matrix presented as $\mathbf{\Gamma} = \mathbf{p}\tilde{\mathbf{C}}\mathbf{p}$, where $\mathbf{p}$ is the slowness vector, are also all positive,

$$\lambda = \begin{bmatrix} 1 & \dfrac{v_{\text{phs},S1}^2}{v_{\text{phs},P}^2} & \dfrac{v_{\text{phs},S2}^2}{v_{\text{phs},P}^2} \end{bmatrix} \quad . \tag{C7}$$

Note that for both representations of the Christoffel matrix, $\hat{\mathbf{\Gamma}} = \mathbf{n}\tilde{\mathbf{C}}\mathbf{n}$ and $\mathbf{\Gamma} = \mathbf{p}\tilde{\mathbf{C}}\mathbf{p}$, the eigenvectors are the polarizations.

The phase velocity (or slowness) direction can be arbitrary. As demonstrated by Grechka (2020), in a complex triclinic medium, some polarization directions may be "prohibited" (never exist). This phenomenon is due to combination of the stiffness components that match definite restriction criteria derived in the referred paper. On the other hand, the slowness directions are not restricted. Consequently, the set of all feasible polarizations is a subset of all slowness directions. Hence, we conclude: Since for any slowness direction, the Christoffel matrix $\hat{\mathbf{\Gamma}} = \mathbf{n}\tilde{\mathbf{C}}\mathbf{n}$ is normally positive definite, then, for any polarization, the matrix $\mathbf{g}\tilde{\mathbf{C}}\mathbf{g}$ is normally also positive definite. This means that the matrix $\mathbf{g}\tilde{\mathbf{C}}\mathbf{g}$ is invertible, because its determinant is the product of the eigenvalues. The term "normally" means a natural non-degenerative stiffness tensor $\tilde{\mathbf{C}}$. The issue arises when applying an acoustic approximation. This case is considered in Appendix I, where an acoustic ellipsoidal medium is studied. In this case, matrix $\mathbf{g}\tilde{\mathbf{C}}\mathbf{g}$ becomes positive semidefinite, with a vanishing determinant, and its inverse does not exist. However, even in this case, the Finsler metric tensor $\mathbf{G}(\mathbf{x},\mathbf{p}) = \left[H_{\mathbf{pp}}^{\lambda}(\mathbf{x},\mathbf{p})\right]^{-1}$ may exist, provided the Hamiltonian can be factorized such that the shear factor is further removed.

A proposed method to compute the Finsler metric $\mathbf{G}$

As mentioned, computing $H_{\mathbf{pp}}^{\lambda}$ needed to establish the Finsler metric is not easy due to the dependence of the eigenvalue Hamiltonian on the polarization vector. We propose an alternative method to compute the Finsler metric $\mathbf{G}$, where $\mathbf{G}^{-1} = H_{\mathbf{pp}}^{\lambda}$ by replacing the time-related eigenvalue Hamiltonian $H^{\lambda}$ (defined in equation 30) by another time-related Hamiltonian $H^{\tau} = \dfrac{H^{\bar{\tau}}}{\mathbf{p} \cdot H_{\mathbf{p}}^{\bar{\tau}}}$ (defined in equation C9 of Part I) which doesn't explicitly depend on the polarization vector $\mathbf{g}(\mathbf{x},\mathbf{p})$. For this Hamiltonian, the Hessian $H_{\mathbf{pp}}^{\tau}$ can be computed analytically in a straightforward way. Then, computing the inverse Finsler metric requires an additional scalar normalization,

$$\mathbf{G}^{-1} = \frac{H_{\mathbf{pp}}^{\tau}(\mathbf{x},\mathbf{p})}{\mathbf{p} \cdot H_{\mathbf{pp}}^{\tau}(\mathbf{x},\mathbf{p}) \cdot \mathbf{p}} \quad , \qquad \mathbf{G}^{-1}\mathbf{p} = \mathbf{v}_{\text{ray}} \quad , \tag{C8}$$

which leads to,

$$\mathbf{p}\mathbf{G}^{-1}\mathbf{p} = 1 \quad , \qquad \mathbf{v}_{\text{ray}} \mathbf{G} \mathbf{v}_{\text{ray}} = 1 \quad . \tag{C9}$$

Note that the last equation of set C9 is equation 57 in the research report by Červený (2002b).

Introduction of equation C9 into the unmodified, first-order homogeneity Lagrangian $L^U$ and changing the flow variable traveltime by the arclength, leads to,

$$t = \int_S^R \sqrt{\mathbf{v}_{\text{ray}} \mathbf{G} \mathbf{v}_{\text{ray}}} \, d\tau = \int_S^R \sqrt{\mathbf{r} \, v_{\text{ray}} \mathbf{G} \mathbf{r} \, v_{\text{ray}}} \, d\tau = \int_S^R \sqrt{\mathbf{r} \mathbf{G} \mathbf{r}} \, v_{\text{ray}} d\tau = \int_S^R \sqrt{\mathbf{r} \mathbf{G} \mathbf{r}} \, ds \quad , \quad (C10)$$

where,

$$\mathbf{r} \mathbf{G} \mathbf{r} = v_{\text{ray}}^{-2} \quad . \quad (C11)$$

Remark: We further note that for the Hamiltonian $H^\lambda$ suggested by Červený (2002a, 2002b) and defined in equation 30, the denominator on the right-hand side of the first equation of set C8 is identically 1, $\mathbf{p} \cdot H^\lambda_{\mathbf{pp}}(\mathbf{x},\mathbf{p}) \cdot \mathbf{p} = 1$, and therefore the normalization is not needed when using the eigenvalue Hamiltonian, $\mathbf{G}^{-1} = H^\lambda_{\mathbf{pp}}(\mathbf{x},\mathbf{p})$. However, computing the Hessian $H^\lambda_{\mathbf{pp}}$ using equation 30 is not trivial. Indeed, we are aware of several challenging ways to compute the second derivatives of the eigenvalue of the Christoffel matrix wrt the location and slowness components (e.g., Červený, 1972; and Gajewski and Pšenčík, 1990). However, in our mind, computing the Hessian $H^\tau_{\mathbf{pp}}$ from the traveltime-scaled Hamiltonian $H^\tau$ (defined in equation C9 of Part I) is simpler, as $H^\tau$ is an explicit function of the slowness vector components and does not include other slowness-dependent objects (like the polarization vector).

# APPENDIX D. INVARIANCE OF KINEMATIC EQUATIONS FROM ALL FORMS OF LAGRANGIANS

In this appendix we prove that the kinematic ray tracing equations, that follow from the proposed Lagrangian (equation 1) and the two alternative Lagrangians suggested by Červený (2002a, 2002b) (equation 29), are identical.

As already mentioned, for the flow variable traveltime, Červený (2002a, 2002b) suggests two alternative forms of the Lagrangian, with a first- and second-degree homogeneity wrt the components of $\dot{\mathbf{x}}_\tau \equiv d\mathbf{x}/d\tau = \mathbf{v}_{\text{ray}}$,

$$L^U(\tau) = \sqrt{\mathbf{v}_{\text{ray}} \mathbf{G} \mathbf{v}_{\text{ray}}} = 1 \quad , \quad L^M(\tau) = \frac{1}{2} \mathbf{v}_{\text{ray}} \mathbf{G} \mathbf{v}_{\text{ray}} = \frac{1}{2} \quad . \tag{D1}$$

The former is the unmodified Lagrangian, and the latter is modified. The modified Lagrangian, $L^M(\mathbf{x}, \dot{\mathbf{x}}_\tau)$, is related to the eigenvalue Hamiltonian $H^\lambda(\mathbf{x}, \mathbf{p})$ via the Legendre transform (equation 31), and the components of the ray velocity can be computed with the use of the momentum equation alone, as $\det L^M_{\dot{\mathbf{x}}_\tau \dot{\mathbf{x}}_\tau} \neq 0$.

In both cases, the Lagrangian is defined for the flow parameter traveltime and is a constant value, $L^U(\tau) = 1$, $L^M(\tau) = 1/2$ (because $\mathbf{G}\mathbf{v}_{\text{ray}} = \mathbf{p}$ and $\mathbf{v}_{\text{ray}} \cdot \mathbf{p} = 1$), and in both cases the generalized momentum of the Lagrangian is the slowness vector,

$$L^U_{\dot{\mathbf{x}}_\tau} = L^M_{\dot{\mathbf{x}}_\tau} = \frac{\partial L^U(\tau)}{\partial \mathbf{v}_{\text{ray}}} = \frac{\partial L^M(\tau)}{\partial \mathbf{v}_{\text{ray}}} = \mathbf{p} = \mathbf{G} \mathbf{v}_{\text{ray}} \quad , \tag{D2}$$

which constitutes the first kinematic ray tracing equation.

To establish the second kinematic ray tracing equation, we note that since the Lagrangian is constant, its full differential vanishes,

$$\frac{\partial L^M\left(\mathbf{x}, \mathbf{v}_{ray}\right)}{\partial \mathbf{v}_{ray}} \cdot \Delta \mathbf{v}_{ray} + \frac{\partial L^M\left(\mathbf{x}, \mathbf{v}_{ray}\right)}{\partial \mathbf{x}} \cdot \Delta \mathbf{x} = 0 \quad , \tag{D3}$$

which can be arranged as,

$$\mathbf{p} \cdot \Delta \mathbf{v}_{ray} + \frac{\partial L^M\left(\mathbf{x}, \mathbf{v}_{ray}\right)}{\partial \mathbf{x}} \cdot \Delta \mathbf{x} = 0 \quad . \tag{D4}$$

Now, consider a particular case where the ray velocity direction $\mathbf{r}$ is fixed, but its magnitude $v_{ray}$ varies due to a change of the position $\Delta \mathbf{x}$. In this case, $\Delta \mathbf{v}_{ray} = \Delta v_{ray} \mathbf{r}$, and the equation above becomes,

$$\mathbf{p} \cdot \mathbf{r} \Delta v_{ray} + \frac{\partial L^M\left(\mathbf{x}, \mathbf{v}_{ray}\right)}{\partial \mathbf{x}} \cdot \Delta \mathbf{x} = 0 \quad . \tag{D5}$$

Recall that $\mathbf{p} \cdot \mathbf{r} = v_{ray}^{-1}$, and this leads to,

$$\frac{\Delta v_{ray}}{v_{ray}} = -\frac{\partial L^M\left(\mathbf{x}, \mathbf{v}_{ray}\right)}{\partial \mathbf{x}} \cdot \Delta \mathbf{x} \tag{D6}$$

We can consider separately three cases where only $\Delta x_i$ varies, $i = 1, 2, 3$ (one index at a time), and arrange equation D6 as,

$$\frac{\partial L^M}{\partial x_i} = -\frac{1}{v_{ray}} \frac{\Delta v_{ray}}{\Delta x_i} \quad \rightarrow \quad \frac{\partial L^M}{\partial \mathbf{x}} = -\frac{\nabla_{\mathbf{x}} v_{ray}}{v_{ray}} \quad , \tag{D7}$$

where,

$$\frac{d\mathbf{p}}{d\tau} = \frac{\partial L^M}{\partial \mathbf{x}} = -\frac{\partial H^\lambda}{\partial \mathbf{x}} \qquad (D8)$$

For the flow variable arclength, $ds = v_{\text{ray}} d\tau$, equations D7 and D8 leads to the second equation of set 17 of Part I,

$$\frac{d\mathbf{p}}{ds} = -\frac{\nabla_{\mathbf{x}} v_{\text{ray}}}{v_{\text{ray}}^2} , \qquad (D9)$$

which is the second kinematic equation for both, the modified time-related Lagrangian, $L^M(\tau)$, and the proposed arclength-related Lagrangian, $L(s)$. (Recall the momentum equation, $L_{\dot{\mathbf{r}}} = \mathbf{p}$ or $L^M_{\dot{\mathbf{x}}_\tau} = \mathbf{p}$, can be considered the first kinematic equation.)

Note that the same result is obtained when using in the equations of this appendix the non-modified first-degree homogeneous Lagrangian $L^U(\tau)$ suggested by Červený (2002a, 2002b) instead of the second-degree homogeneous $L^M(\tau)$. Thus, despite the differences between the proposed Lagrangian $L(s)$ and both types of Lagrangian suggested by Červený (2002a, 2002b), $L^U(\tau)$ and $L^M(\tau)$, all three Lagrangians yield the same kinematic ray tracing equations.

Comment. The first- and second-degree homogeneous Lagrangians, $L^U(\tau)$ and $L^M(\tau)$ listed in equation D1, depend on the Finsler metric $\mathbf{G}(\mathbf{x}, \mathbf{v}_{\text{ray}})$ which, in turn, is given by $\left(H^\lambda_{\mathbf{pp}}\right)^{-1}$ (equation A2). However, in the proof that the slowness derivative $d\mathbf{p}/ds$ is invariant whether we use our arclength-related Lagrangian (defined in equation 1) or the Lagrangians suggested by Červený (2002a, 2002b), we never used explicitly the specific form of the Finsler metric. In this proof (equations D3-D8) we only accounted for the established facts that a) for the flow variable traveltime, the Lagrangian is constant along the path (e.g., 1 for $L^U(\tau)$ and 1/2 for $L^M(\tau)$; the specific value of the constant is not essential), and b) the generalized momentum of all three Lagrangians is the slowness vector, $\partial L / \partial \dot{\mathbf{x}} = \mathbf{p}$, $\partial L^M / \partial \dot{\mathbf{x}}_\tau = \mathbf{p}$ and $\partial L^U / \partial \dot{\mathbf{x}}_\tau = \mathbf{p}$ (where $\dot{\mathbf{x}}_\tau = d\mathbf{x}/d\tau = \mathbf{v}_{\text{ray}}$ and $\dot{\mathbf{x}} = d\mathbf{x}/ds = \mathbf{r}$).

# APPENDIX E.

## LAGRANGIAN/HAMILTONIAN TEST FOR TILTED ORTHORHOMBIC MEDIUM

Consider a tilted orthorhombic medium (TOR) with the following elastic properties at a reference node,

$$v_{P,\text{o}} = 3.5 \text{ km/s} \;, \quad f = \frac{v_P^2 - v_{S1}^2}{v_P^2} = \frac{C_{33} - C_{55}}{C_{33}} = 0.75 \;, \tag{E1}$$

where $v_{S1}$ is the shear wave velocity along the "crystal" (local, tilted) axis $x_3$ and polarized along the crystal axis $x_1$. The Tsvankin (1997) orthorhombic parameters are,

$$\delta_1 = -0.05, \; \delta_2 = 0.15, \; \delta_3 = 0.08, \; \varepsilon_1 = 0.12, \; \varepsilon_2 = 0.29, \; \gamma_1 = -0.08, \; \gamma_2 = 0.10 \;. \tag{E2}$$

The orientation of the crystal frame wrt the global frame can be defined by three successive rotations with Euler's angles: azimuth $\psi_{\text{rot}}$, zenith $\theta_{\text{rot}}$ and spin $\xi_{\text{rot}}$. The sequence of "global to local" rotation of the coordinate frame is as follows (see details in Ravve and Koren, 2019):

- Rotation about global axis $z$ for azimuth $\psi_{rot}$

- Rotation about axis $y_v$ of the first intermediate frame for zenith (tilt) $\theta_{rot}$

- Rotation about axis $z_w$ of the second intermediate frame for spin $\xi_{rot}$

Assume the following values of zenith $\theta_{rot}$, azimuth $\psi_{rot}$ and spin $\xi_{rot}$,

$$\theta_{rot} = 32^o = 0.55850536 \text{ rad},$$
$$\psi_{rot} = 112^o = 1.95476876 \text{ rad},$$
$$\xi_{rot} = 69^o = 1.20427718 \text{ rad} \quad . \quad (E3)$$

The "global to local" rotational matrix reads,

$$\mathbf{A}_{rot} = \begin{bmatrix} -0.97944861 & -0.067941918 & -0.18990608 \\ -0.035689031 & -0.86831802 & +0.49472225 \\ -0.19851125 & +0.49133529 & +0.84804810 \end{bmatrix} \quad . \quad (E4)$$

The computed density-normalized components of the stiffness tensor in the orthorhombic crystal (tilted) frame are ($km^2/s^2$),

$$\begin{array}{lll} C_{11} = 19.355 & C_{22} = 15.19 & C_{33} = 12.25 \\ C_{12} = 15.692887 & C_{13} = 7.8082966 & C_{23} = 7.3302213 \\ C_{44} = 2.14375 & C_{55} = 3.0625 & C_{66} = 2.5725 \end{array} \quad . \quad (E5)$$

Depending on the ray tracing application, there is a tradeoff whether to perform the computations in the global frame or in the spatially varying tilted (local or "crystal") frames. In the first approach, the orthorhombic stiffness tensor at each subsurface grid point is rotated once to the global frame, resulting in "populating" all 21 stiffness parameters at each grid point. Hence, with this approach, the ray tracing application requires operating over all 21 stiffness components which need to be kept in the memory of the computer. The second approach allows for operating directly over the orthorhombic parameters, but requires a huge amount of rotations of the vectors from local to global frame and vice versa.

In this example we choose to adopt the first approach where the computations are directly performed in the global frame. Applying the rotation technique suggested by Bond (1943), we obtain the (symmetric) Kelvin-form stiffness tensor in the global frame,

$$\mathbf{C}_{\text{glb}} = \begin{bmatrix} +18.936887 & +13.513217 & +9.7906361 & -4.5262926 & +1.2238081 & +0.78622210 \\ +13.513217 & +13.754844 & +8.3533406 & -1.4281954 & +1.6278592 & +0.80757635 \\ +9.7906361 & +8.3533406 & +12.451692 & -0.33934785 & +0.94679060 & +0.33625404 \\ -4.5262926 & -1.4281954 & -0.33934785 & +5.7095620 & -1.0994471 & -0.35360136 \\ +1.2238081 & +1.6278592 & +0.94679060 & -1.0994471 & +6.0422293 & +0.58916968 \\ +0.78622210 & +0.80757635 & +0.33625404 & -0.35360136 & +0.58916968 & +5.4572858 \end{bmatrix}.$$

(E6)

Assume a compressional wave with the following components of the ray velocity direction in the global frame,

$$\mathbf{r} = \begin{bmatrix} 0.8560 & 0.4992 & 0.1344 \end{bmatrix} \quad . \tag{E7}$$

We start by solving equation set 18 of Part I for the unknown slowness components. Assuming that the given ray direction is close to the slowness direction, we solve the Christoffel equation for the initial guess slowness components (s/km),

$$\begin{aligned} p_1^{\text{ini}} &= 0.19487831 \quad , \quad p_2^{\text{ini}} = 0.11364866 \, , \\ p_3^{\text{ini}} &= 0.030597716 \, , \quad v_{\text{phs}} = 4.3924847 \quad . \end{aligned}$$

(E8)

The final solution for the compressional slowness is (s/km),

$$p_1 = 0.18215367, \quad p_2 = 0.13015075, \quad p_3 = 0.06477595 \quad . \tag{E9}$$

The phase and ray velocities are,

$$v_{\text{phs}} = 4.2908142 \text{ km/s} \quad v_{\text{ray}} = 4.3553878 \text{ km/s} \quad . \tag{E10}$$

Now assume that the tilted (crystal) axial (local $x_3$) compressional velocity $v_P(\mathbf{x})$ changes linearly in space, while all other (relative, unitless) model parameters are kept constant (including the orientation angles of the symmetry planes). This type of model is referred to by Červený (2000), Section 3.6.6, as a factorized anisotropic inhomogeneous medium (FAI). For simplicity, the reference node is located at the origin of the Cartesian frame, so that,

$$v_P = v_{P,\text{o}} + \nabla v_P \cdot \mathbf{x} \quad . \tag{E11}$$

The parameter values in equations E1 and E2 are related to the origin of the reference frame. The gradient $\nabla v_P$ is constant, but not collinear with the tilted crystal $x_3$ axis. In this example, we set the gradient in the global frame, to be,

$$\nabla v_P = k_m \begin{bmatrix} -0.224 & 0.600 & 0.768 \end{bmatrix}, \quad k_m = 1\text{s}^{-1} \quad , \tag{E12}$$

where the vector in the brackets is the normalized direction of the gradient, and $k_m$ is the gradient magnitude. The orientation of the tilted axis in the global frame is given by the third row of the rotation matrix $\mathbf{A}_{\text{rot}}^{(3)}$. The angle between this axis and the velocity gradient direction reads,

$$\alpha_{\nabla v,\text{ax}} = \arccos \frac{\mathbf{A}^{(3)} \cdot \nabla v_P}{k_m} = 0.137462 \text{ rad} = 7.87597^\circ \quad . \tag{E13}$$

In this simple example, with all parameters of the TOR medium constant, except the axial compressional velocity, this velocity becomes a scaling factor. We can therefore compute the ray velocity for $v_P = 1$, and then rescale it for any other value of $v_P$ (in particular, for $v_P$ of the neighboring points). As a result, the spatial gradient of the ray velocity becomes collinear with the gradient of the axial velocity

$$\nabla v_{\text{ray}} = \nabla v_P \frac{v_{\text{ray},o}}{v_{P,o}} \quad , \tag{E14}$$

where $v_{P,o}$ is given in equation E1, and $v_{\text{ray},o}$ in equation E10. The slowness derivative obtained with the Lagrangian becomes,

$$\left( \frac{d\mathbf{p}}{ds} \right)_L = -\frac{\nabla v_P}{v_{P,o} v_{\text{ray},o}} \quad . \tag{E15}$$

The subscript $L$ means "computed with the use of the Lagrangian". The derivative of the slowness wrt the arclength, computed with the Lagrangian, becomes,

$$\left( \frac{d\mathbf{p}}{ds} \right)_L = [+1.4694443 \quad -3.9360116 \quad -5.0380949] \cdot 10^{-2} \text{ s/km}^2 \quad . \tag{E16}$$

Next, we compute the slowness derivative with the Hamiltonian. For this we need two gradients of the Hamiltonian: wrt the location and wrt the slowness. Both gradients can be computed analytically. We apply the reference Hamiltonian, $H^{\bar{\tau}} = \det(\mathbf{p}\tilde{\mathbf{C}}\mathbf{p} - \mathbf{I})$. The spatial gradient of the reference Hamiltonian reads,

$$H^{\bar{\tau}}_{\mathbf{x}} = \begin{bmatrix} 9.5885694 \cdot 10^{-2} & 0.25683668 & 0.32875095 \end{bmatrix} \text{ km}^{-1} \quad . \tag{E17}$$

The slowness gradient of the reference Hamiltonian reads,

$$H_{\mathbf{p}}^{\bar{\tau}} = [5.5856592 \quad 3.2574312 \quad 0.8770070] \text{ km/s} \quad . \quad (E18)$$

The derivative of the slowness wrt the arclength, computed with our proposed arclength-related Hamiltonian $H(\mathbf{x},\mathbf{p})$, becomes,

$$\left(\frac{d\mathbf{p}}{ds}\right)_H = -H_{\mathbf{x}} = -\frac{H_{\mathbf{x}}^{\bar{\tau}}}{\sqrt{H_{\mathbf{p}}^{\bar{\tau}} \cdot H_{\mathbf{p}}^{\bar{\tau}}}} = [1.4694443 \quad -3.9360116 \quad -5.0380949] \cdot 10^{-2} \text{ s/km}^2 \quad . \quad (E19)$$

The subscript $H$ at $d\mathbf{p}/ds$ on the left-hand side of equation E19 means that this vector has been computed with the use of the arclength-related Hamiltonian. The same result can be obtained with the eigenvalue Hamiltonian $H^\lambda$ suggested by Červený (2000, 2002a, 2002b). For this we first find the polarization at the reference node,

$$\mathbf{g} = [0.83685224 \quad 0.51752874 \quad 0.17844420] \quad , \quad (E20)$$

and then we apply equation B15,

$$\left(\frac{d\mathbf{p}}{ds}\right)_{H^\lambda} = -\frac{\mathbf{g}\left(\mathbf{p}\nabla_{\mathbf{x}}^T \tilde{\mathbf{C}} \mathbf{p}\right)\mathbf{g}}{2v_{\text{ray}}} = [1.4694443 \quad -3.9360116 \quad -5.0380949] \cdot 10^{-2} \text{ s/km}^2 \quad (E21)$$

As we see, the values of the three derivatives (equations E16, E19 and E21) are identical up to eight digits. In fact, the accuracy is even better. The relative error reads,

$$E_{\dot{\mathbf{p}}} = \frac{\left|(d\mathbf{p}/ds)_L - (d\mathbf{p}/ds)_H\right|}{\left|(d\mathbf{p}/ds)_H\right|} = 1.832 \cdot 10^{-16} \quad \text{(proposed Lagrangian)} \quad,$$

$$E_{\dot{\mathbf{p}}} = \frac{\left|(d\mathbf{p}/ds)_{L^M} - (d\mathbf{p}/ds)_{H^\lambda}\right|}{\left|(d\mathbf{p}/ds)_{H^\lambda}\right|} = 1.090 \cdot 10^{-16} \quad \text{(Červený modified Lagrangian)} \quad. \tag{E22}$$

We also make sure that,

$$H_{\mathbf{p}} = \frac{H_{\mathbf{p}}^{\bar{\tau}}}{\sqrt{H_{\mathbf{p}}^{\bar{\tau}} \cdot H_{\mathbf{p}}^{\bar{\tau}}}} = [0.8560 \quad 0.4992 \quad 0.1344] = \mathbf{r} = \frac{d\mathbf{x}}{ds} \quad . \tag{E23}$$

The ray velocity direction computed with equation B10 yields the same result.

We compute the symmetric positive-definite (and thus, invertible) tensor $\mathbf{F}$ that relates the ray velocity vector to the slowness vector,

$$\mathbf{F}^{-1} = \left(\mathbf{g}\tilde{\mathbf{C}}\mathbf{g}\right)^{-1} = \begin{bmatrix} +0.16515671 & -0.20645400 & +0.026118295 \\ -0.20645400 & +0.44551724 & -0.11752142 \\ +0.026118295 & -0.11752142 & +0.38081870 \end{bmatrix} \quad (\text{km/s})^{-2} \quad . \tag{E24}$$

Finally, we apply equation C8 to compute the Finsler metric from Hamiltonian $H^\tau(\mathbf{x},\mathbf{p})$ defined in equation C9 of Part I,

$$\mathbf{G} = \begin{bmatrix} -4.9694700 \cdot 10^{-3} & +8.1151705 \cdot 10^{-2} & +4.1410271 \cdot 10^{-2} \\ +8.1151705 \cdot 10^{-2} & -7.7692100 \cdot 10^{-2} & -5.9468965 \cdot 10^{-3} \\ +4.1410271 \cdot 10^{-2} & -5.9468965 \cdot 10^{-3} & -1.3099627 \cdot 10^{-1} \end{bmatrix} \quad (\text{km/s})^{-2} \quad . \tag{E25}$$

The slowness components are given in equation E9, the ray velocity direction in equation E7, and its magnitude in equation E10. With tensor $\mathbf{F}$ and with tensor $\mathbf{G}$, we test independently the two-way transform of equations C2 and C4 between the ray velocity vector components and the slowness vector components.

**APPENDIX F. LAGRANGIAN/HAMILTONIAN TEST FOR TRICLINIC MEDIUM**

In this section, we carry out a similar test, but for a triclinic medium with all 21 coefficients varying independently (not a FAI medium). An example of the stiffness tensor for a medium with triclinic symmetry is given by Grechka (2017).

All 21 stiffness components are assumed varying in space. Locally, in an infinitesimal proximity of the reference node, this dependence can be considered linear or linearized. In Table 1, we provide the values of the density-normalized stiffness tensor components and their relative (normalized) gradients,

$$\overline{\nabla_{\mathbf{x}} C_{ij}} = \frac{\nabla_{\mathbf{x}} C_{ij}}{C_{ij}} \quad . \tag{F1}$$

For the gradient components of the stiffness, we applied random numbers within the range,

$$-0.1 \text{ km}^{-1} \leq \overline{\nabla_{\mathbf{x}} C_{ij,k}} \leq +0.5 \text{ km}^{-1} \quad , \quad k = 1, 2, 3 \quad . \tag{F2}$$

Here $i, j$ are indices of the stiffness matrix, and $k$ is the index of its Cartesian gradient component. The range is asymmetric, biased to the positive side. When the range of the gradient components is symmetric, the effect of positive and negative changes of the stiffness components on the ray velocity is partially compensated, and the spatial gradient of the ray velocity becomes relatively small. Therefore, we apply the asymmetric range.

Assume the following components of the ray velocity direction,

$$\mathbf{r} = \begin{bmatrix} 0.224 & 0.600 & 0.768 \end{bmatrix} \quad . \tag{F3}$$

Compute the slowness with equation D4 of Part I, assuming the weight of the Hamiltonian term in the target function, $w = 45 \text{ km}^2/\text{s}^2$. This value does not affect the solution, but extremely small or large values may reduce the accuracy. The initial guess for the compressional slowness vector reads,

$$\begin{aligned} p_1^{\text{ini}} &= 4.3678631 \cdot 10^{-2} \text{ s/km} \;, & p_2^{\text{ini}} &= 1.1699633 \cdot 10^{-1} \text{ s/km} \;, \\ p_3^{\text{ini}} &= 1.4975531 \cdot 10^{-1} \text{ s/km} \;, & v_{\text{phs}} &= 5.1283659 \quad \text{km/s.} \end{aligned} \tag{F4}$$

The final solution reads (s/km),

$$\begin{aligned} p_1 &= 1.2893238 \cdot 10^{-1} \;, \\ p_2 &= 2.5360427 \cdot 10^{-1} \;, \end{aligned} \qquad p_3 = 1.4179428 \cdot 10^{-1} \quad . \tag{F5}$$

The phase and ray velocities of the compressional wave are,

$$v_{\text{phs}} = 3.1458940 \text{ km/s} \qquad v_{\text{ray}} = 3.4489725 \text{ km/s} \quad . \tag{F6}$$

Next, we recall that the stiffness components are space-dependent,

$$C_{ij}(\mathbf{x}) = C_{ij}^o + \nabla_{\mathbf{x}} C_{ij} \cdot \mathbf{x} = C_{ij}^o \left(1 + \overline{\nabla_{\mathbf{x}} C_{ij}} \cdot \mathbf{x}\right) \quad , \tag{F7}$$

where the upper index "o" means "related to the reference node located at the origin of the frame". The spatial gradient of the ray velocity can be computed analytically (Ravve and Koren, 2019). First, we compute the spatial gradient of the slowness vector, $\mathbf{p}_{\mathbf{x}} = \partial \mathbf{p}/\partial \mathbf{x}$, which is a $3\times 3$ matrix (tensor). For this, we assume the weight $w = 45 \text{ km}^2/\text{s}^2$ build two $3\times 3$ matrices, $f_{\mathbf{pp}}$ and $f_{\mathbf{px}}$, defined by equations D9 and D11 of Part I,

$$f_{\mathbf{pp}} = \begin{bmatrix} +7.7623844\cdot 10^2 & +1.0621595\cdot 10^3 & -1.0430900\cdot 10^3 \\ +1.0621595\cdot 10^3 & +6.2389341\cdot 10^3 & -5.1488075\cdot 10^3 \\ -1.0430900\cdot 10^3 & -5.1488075\cdot 10^3 & +4.3717404\cdot 10^3 \end{bmatrix} , \quad \text{(F8)}$$

and,

$$f_{\mathbf{px}} = \begin{bmatrix} +5.4014887 & -2.7183728 & +51.328241 \\ -73.022039 & -27.767468 & +251.98913 \\ +55.790573 & +23.924545 & -210.22332 \end{bmatrix} . \quad \text{(F9)}$$

Next, we apply equation E24 of Part I to compute the spatial gradient of the slowness vector, $\mathbf{p}_\mathbf{x} = \partial \mathbf{p}/\partial \mathbf{x}$,

$$\mathbf{p}_\mathbf{x} = \begin{bmatrix} -3.1846835\cdot 10^{-2} & -9.1534896\cdot 10^{-3} & -1.4496588\cdot 10^{-2} \\ +1.1532673\cdot 10^{-2} & -1.1044963\cdot 10^{-2} & -3.8930507\cdot 10^{-2} \\ -6.7776606\cdot 10^{-3} & -2.0664733\cdot 10^{-2} & -1.2223077\cdot 10^{-3} \end{bmatrix} \frac{\text{s}}{\text{km}^2} . \quad \text{(F10)}$$

Note that matrices $f_{\mathbf{pp}}$ and $f_{\mathbf{px}}$ depend on our choice of the target function; in particular they depend on the weight factor, $w$. Contrarily, the spatial slowness gradient, $\mathbf{p}_\mathbf{x}$, represents a physical object independent of this choice. Note that this matrix is not symmetric. This is not the matrix of the second derivatives of the traveltime wrt the coordinates of the arrival point, $\mathbf{p}_\mathbf{x} = \partial \mathbf{p}/\partial \mathbf{x} \neq d^2\tau/d\mathbf{x}^2$, because for the computation of $\mathbf{p}_\mathbf{x}$ the ray direction $\mathbf{r}$ is assumed constant. We recall that due to the momentum equation, the slowness vector is the directional gradient of the Lagrangian, $\mathbf{p} = L_{\mathbf{r}}$, and therefore, the spatial slowness gradient is equal to the mixed Hessian of the Lagrangian, $\mathbf{p_x} = L_{\mathbf{rx}}$. The spatial slowness gradient was also computed directly from the Hamiltonian Hessians, $H_{\mathbf{pp}}$ and $H_{\mathbf{px}}$, with equation E25 of Part I; the results are identical to equation F10.

Finally, we apply equation E23 of Part I to compute the spatial gradient of the ray velocity,

$$\nabla_{\mathbf{x}} v_{\text{ray}} = \begin{bmatrix} +6.4465165 \cdot 10^{-2} & +0.29200707 & +0.32765045 \end{bmatrix} \text{ s}^{-1} \quad , \tag{F11}$$

where the absolute value of this gradient reads, $\left| \nabla_{\mathbf{x}} v_{\text{ray}} \right| = 0.4436 \text{ s}^{-1}$.

Next, we compute the gradients of the Hamiltonian. The spatial gradient of the reference Hamiltonian reads,

$$H_{\mathbf{x}}^{\bar{\tau}} = \begin{bmatrix} +9.0553433 \cdot 10^{-4} & +4.1017878 \cdot 10^{-3} & +4.6024661 \cdot 10^{-3} \end{bmatrix} \text{ km}^{-1} \quad . \tag{F12}$$

The slowness gradient of the reference Hamiltonian $H^{\bar{\tau}}$ reads,

$$H_{\mathbf{p}}^{\bar{\tau}} = \begin{bmatrix} +3.7428921 \cdot 10^{-2} & +1.0025604 \cdot 10^{-1} & +1.2832773 \cdot 10^{-1} \end{bmatrix} \text{ km/s} \quad . \tag{F13}$$

The derivative of the slowness wrt the arclength, computed with the arclength-related Lagrangian, becomes,

$$\left( \frac{d\mathbf{p}}{ds} \right)_L = -\frac{\nabla_{\mathbf{x}} v_{\text{ray}}}{v_{\text{ray}}^2} = \begin{bmatrix} -0.54193304 & -2.4547875 & -2.7544272 \end{bmatrix} \cdot 10^{-2} \text{ s/km}^2 \quad . \tag{F14}$$

Next, we compute the same derivative with the arclength-related Hamiltonian, $H(\mathbf{x},\mathbf{p})$, and we obtain,

$$\left(\frac{d\mathbf{p}}{ds}\right)_H = -\frac{H_{\mathbf{x}}^{\bar{\tau}}}{\sqrt{H_{\mathbf{p}}^{\bar{\tau}} \cdot H_{\mathbf{p}}^{\bar{\tau}}}} = -H_{\mathbf{x}} = [-0.54193304 \quad -2.4547875 \quad -2.7544272] \cdot 10^{-2} \text{ s/km}^2 \;. \quad \text{(F15)}$$

To compute the slowness derivative wrt arclength using the eigenvalue Hamiltonian suggested by Červený (2000, 2002a, 2002b), we first compute the polarization vector, corresponding to the Christoffel matrix eigenvalue 1,

$$\mathbf{g} = [0.30818933 \quad 0.72063993 \quad 0.62104544] \;, \quad \text{(F16)}$$

and then apply equation B15,

$$\left(\frac{d\mathbf{p}}{ds}\right)_{H^\lambda} = -\frac{\mathbf{g}\left(\mathbf{p}\nabla_{\mathbf{x}}^T \tilde{\mathbf{C}} \mathbf{p}\right)\mathbf{g}}{2 v_{\text{ray}}} = [-0.54193304 \quad -2.4547875 \quad -2.7544272] \cdot 10^{-2} \text{ s/km}^2 \;. \quad \text{(F17)}$$

As we can see, in equations F14, F15 and F17, the first eight digits coincide. Note that the accuracy is even better. The relative errors are,

$$E_{\dot{\mathbf{p}}} = \frac{|(d\mathbf{p}/ds)_L - (d\mathbf{p}/ds)_H|}{|(d\mathbf{p}/ds)_H|} = \left\{ \underbrace{2.344 \cdot 10^{-14}}_{\text{our Hamiltonian}} \quad \underbrace{3.239 \cdot 10^{-16}}_{\text{by Červený}} \right\} \;. \quad \text{(F18)}$$

We also make sure that the arclength-related Hamiltonian yields the correct ray velocity direction,

$$H_{\mathbf{p}} = \frac{H_{\mathbf{p}}^{\bar{\tau}}}{\sqrt{H_{\mathbf{p}}^{\bar{\tau}} \cdot H_{\mathbf{p}}^{\bar{\tau}}}} = [0.224 \quad 0.600 \quad 0.768] = \mathbf{r} = \frac{d\mathbf{x}}{ds} \quad .\quad (F19)$$

The same result for the ray velocity direction can be obtained with equation B10.

Next, we compute tensor $\mathbf{F}$ (symmetric, positive-definite, invertible) for the given triclinic medium and the computed polarization,

$$\mathbf{F}^{-1} = (\mathbf{g}\tilde{\mathbf{C}}\mathbf{g})^{-1} = \begin{bmatrix} +2.0689316 \cdot 10^{-1} & -1.0250177 \cdot 10^{-2} & -3.6603192 \cdot 10^{-3} \\ -1.0250177 \cdot 10^{-2} & +1.3174179 \cdot 10^{-1} & -4.1909551 \cdot 10^{-3} \\ -3.6603192 \cdot 10^{-3} & -4.1909551 \cdot 10^{-3} & +5.7873074 \cdot 10^{-2} \end{bmatrix} \quad . \quad (F20)$$

Finally, we apply equation C8 to compute the Finsler metric,

$$\mathbf{G} = \begin{bmatrix} +4.9654381 \cdot 10^{-1} & -8.0404455 \cdot 10^{-2} & -3.3333728 \cdot 10^{-2} \\ -8.0404455 \cdot 10^{-2} & +1.6664307 \cdot 10^{-1} & -1.0995910 \cdot 10^{-2} \\ -3.3333728 \cdot 10^{-2} & -1.0995910 \cdot 10^{-2} & +7.1844189 \cdot 10^{-2} \end{bmatrix} (\text{km/s})^{-2} \quad . \quad (F21)$$

The slowness components are given in equation F5, the ray velocity direction in equation F3, and its magnitude in equation F6. With tensors $\mathbf{F}$ and $\mathbf{G}$, we test again (validate numerically) the two-way transform of equations C2 and C4 between the ray velocity vector and the slowness vector.

Completing this numerical example, we provide the gradients and Hessians of the slowness vector and those of the ray velocity magnitude, making this example a useful benchmark for testing the theory in general anisotropic media. The ray direction is given in equation F3, the slowness vector components in equation F5, the ray and phase velocities in equation F6. The spatial gradient of the slowness vector is given in equation F10. Its directional gradient reads,

$$\mathbf{p_r} = \begin{bmatrix} +3.8643656 \cdot 10^{-2} & -9.1071062 \cdot 10^{-3} & -4.1561396 \cdot 10^{-3} \\ -9.1071062 \cdot 10^{-3} & +8.2395475 \cdot 10^{-3} & -3.7809072 \cdot 10^{-3} \\ -4.1561396 \cdot 10^{-3} & -3.7809072 \cdot 10^{-3} & +4.1660411 \cdot 10^{-3} \end{bmatrix} \frac{s}{km} \quad . \quad (F22)$$

The spatial Hessian of the slowness vector is,

$$\mathbf{p_{xx}} = \begin{bmatrix} \begin{pmatrix} +2.1168640 \cdot 10^{-2} \\ +5.8401766 \cdot 10^{-3} \\ +1.0732451 \cdot 10^{-2} \end{pmatrix} & \begin{pmatrix} +5.8401766 \cdot 10^{-3} \\ +1.6070661 \cdot 10^{-3} \\ +3.4471860 \cdot 10^{-3} \end{pmatrix} & \begin{pmatrix} +1.0732451 \cdot 10^{-2} \\ +3.4471860 \cdot 10^{-3} \\ +4.5603531 \cdot 10^{-3} \end{pmatrix} \\ \begin{pmatrix} +4.8098478 \cdot 10^{-4} \\ -1.2119887 \cdot 10^{-3} \\ -4.9029026 \cdot 10^{-3} \end{pmatrix} & \begin{pmatrix} -1.2119887 \cdot 10^{-3} \\ +1.7364137 \cdot 10^{-3} \\ +4.9726737 \cdot 10^{-3} \end{pmatrix} & \begin{pmatrix} -4.9029026 \cdot 10^{-3} \\ +4.9726737 \cdot 10^{-3} \\ +1.7649794 \cdot 10^{-2} \end{pmatrix} \\ \begin{pmatrix} +5.1900638 \cdot 10^{-4} \\ +2.7682447 \cdot 10^{-3} \\ +2.7993052 \cdot 10^{-4} \end{pmatrix} & \begin{pmatrix} +2.7682447 \cdot 10^{-3} \\ +8.3691436 \cdot 10^{-3} \\ +1.2058275 \cdot 10^{-3} \end{pmatrix} & \begin{pmatrix} +2.7993052 \cdot 10^{-4} \\ +1.2058275 \cdot 10^{-3} \\ -5.2835094 \cdot 10^{-4} \end{pmatrix} \end{bmatrix} \frac{s}{km^3} \quad . \quad (F23)$$

The directional Hessian of the slowness vector is,

$$\mathbf{p_{rr}} = \begin{bmatrix} \begin{pmatrix} -2.8142436\cdot 10^{-1} \\ +2.4545894\cdot 10^{-2} \\ +1.2588367\cdot 10^{-2} \end{pmatrix} & \begin{pmatrix} +2.4545894\cdot 10^{-2} \\ +7.3331752\cdot 10^{-3} \\ -1.0300509\cdot 10^{-3} \end{pmatrix} & \begin{pmatrix} +1.2588367\cdot 10^{-2} \\ -1.0300509\cdot 10^{-3} \\ +2.5447604\cdot 10^{-3} \end{pmatrix} \\ \begin{pmatrix} +2.4545894\cdot 10^{-2} \\ +7.3331752\cdot 10^{-3} \\ -1.0300509\cdot 10^{-3} \end{pmatrix} & \begin{pmatrix} +7.3331752\cdot 10^{-3} \\ -2.1006413\cdot 10^{-2} \\ +3.5438400\cdot 10^{-3} \end{pmatrix} & \begin{pmatrix} -1.0300509\cdot 10^{-3} \\ +3.5438400\cdot 10^{-3} \\ +2.4548628\cdot 10^{-3} \end{pmatrix} \\ \begin{pmatrix} +1.2588367\cdot 10^{-2} \\ -1.0300509\cdot 10^{-3} \\ +2.5447604\cdot 10^{-3} \end{pmatrix} & \begin{pmatrix} -1.0300509\cdot 10^{-3} \\ +3.5438400\cdot 10^{-3} \\ +2.4548628\cdot 10^{-3} \end{pmatrix} & \begin{pmatrix} +2.5447604\cdot 10^{-3} \\ +2.4548628\cdot 10^{-3} \\ -8.0846161\cdot 10^{-3} \end{pmatrix} \end{bmatrix} \frac{\text{s}}{\text{km}} \quad . \quad \text{(F24)}$$

The mixed Hessians of the slowness vector are,

$$\mathbf{p_{xr}} = \begin{bmatrix} \begin{pmatrix} -6.5153898\cdot 10^{-3} \\ +9.8249046\cdot 10^{-4} \\ +1.1327513\cdot 10^{-3} \end{pmatrix} & \begin{pmatrix} +2.8585518\cdot 10^{-4} \\ -8.5034414\cdot 10^{-4} \\ +5.8095693\cdot 10^{-4} \end{pmatrix} & \begin{pmatrix} -3.3495909\cdot 10^{-3} \\ +1.1006200\cdot 10^{-3} \\ +1.1710463\cdot 10^{-4} \end{pmatrix} \\ \begin{pmatrix} +9.8249046\cdot 10^{-4} \\ -4.3678174\cdot 10^{-4} \\ +5.4676015\cdot 10^{-5} \end{pmatrix} & \begin{pmatrix} -8.5034414\cdot 10^{-4} \\ +6.4552178\cdot 10^{-4} \\ -2.5629685\cdot 10^{-4} \end{pmatrix} & \begin{pmatrix} +1.1006200\cdot 10^{-3} \\ -6.8844748\cdot 10^{-4} \\ +2.1683543\cdot 10^{-4} \end{pmatrix} \\ \begin{pmatrix} +1.1327513\cdot 10^{-3} \\ +5.4676015\cdot 10^{-5} \\ -3.7310144\cdot 10^{-4} \end{pmatrix} & \begin{pmatrix} +5.8095693\cdot 10^{-4} \\ -2.5629685\cdot 10^{-4} \\ +3.0786143\cdot 10^{-5} \end{pmatrix} & \begin{pmatrix} +1.1710463\cdot 10^{-4} \\ +2.1683543\cdot 10^{-4} \\ -2.0355820\cdot 10^{-4} \end{pmatrix} \end{bmatrix} \frac{\text{s}}{\text{km}^2} \quad , \quad \text{(F25)}$$

and,

$$\mathbf{p_{rx}} = \begin{bmatrix} \begin{pmatrix} -6.5153898 \cdot 10^{-3} \\ +2.8585518 \cdot 10^{-4} \\ -3.3495909 \cdot 10^{-3} \end{pmatrix} & \begin{pmatrix} +9.8249046 \cdot 10^{-4} \\ -8.5034414 \cdot 10^{-4} \\ +1.1006200 \cdot 10^{-3} \end{pmatrix} & \begin{pmatrix} +1.1327513 \cdot 10^{-3} \\ +5.8095693 \cdot 10^{-4} \\ +1.1710463 \cdot 10^{-4} \end{pmatrix} \\ \begin{pmatrix} +9.8249046 \cdot 10^{-4} \\ -8.5034414 \cdot 10^{-4} \\ +1.1006200 \cdot 10^{-3} \end{pmatrix} & \begin{pmatrix} -4.3678174 \cdot 10^{-4} \\ +6.4552178 \cdot 10^{-4} \\ -6.8844748 \cdot 10^{-4} \end{pmatrix} & \begin{pmatrix} +5.4676015 \cdot 10^{-5} \\ -2.5629685 \cdot 10^{-4} \\ +2.1683543 \cdot 10^{-4} \end{pmatrix} \\ \begin{pmatrix} +1.1327513 \cdot 10^{-3} \\ +5.8095693 \cdot 10^{-4} \\ +1.1710463 \cdot 10^{-4} \end{pmatrix} & \begin{pmatrix} +5.4676015 \cdot 10^{-5} \\ -2.5629685 \cdot 10^{-4} \\ +2.1683543 \cdot 10^{-4} \end{pmatrix} & \begin{pmatrix} -3.7310144 \cdot 10^{-4} \\ +3.0786143 \cdot 10^{-5} \\ -2.0355820 \cdot 10^{-4} \end{pmatrix} \end{bmatrix} \frac{\text{s}}{\text{km}^2} \quad . \quad \text{(F26)}$$

The spatial gradient of the ray velocity is listed in equation F11. The directional gradient of the ray velocity reads,

$$\nabla_{\mathbf{r}} v_{\text{ray}} = \begin{bmatrix} -0.76113381 & -0.94734363 & +0.96210957 \end{bmatrix} \text{ km/s} \quad . \quad \text{(F27)}$$

The spatial Hessian of the ray velocity reads,

$$\nabla_{\mathbf{x}} \nabla_{\mathbf{x}} v_{\text{ray}} = \begin{bmatrix} -6.2169897 \cdot 10^{-2} & -2.1285216 \cdot 10^{-2} & +1.6086792 \cdot 10^{-2} \\ -2.1285216 \cdot 10^{-2} & -4.3687616 \cdot 10^{-2} & -2.1151017 \cdot 10^{-4} \\ +1.6086792 \cdot 10^{-2} & -2.1151017 \cdot 10^{-4} & -7.1042263 \cdot 10^{-2} \end{bmatrix} \frac{1}{\text{km} \cdot \text{s}} \quad . \quad \text{(F27)}$$

The directional Hessian of the ray velocity reads,

$$\nabla_{\mathbf{r}} \nabla_{\mathbf{r}} v_{\text{ray}} = \begin{bmatrix} +3.4931631 & +.73180370 & -.59950123 \\ +.73180370 & +3.7665636 & -1.9225502 \\ -.59950123 & -1.9225502 & +.42410003 \end{bmatrix} \frac{\text{km}}{\text{s}} \quad . \quad \text{(F28)}$$

The mixed Hessians of the ray velocity are,

$$\nabla_{\mathbf{x}}\nabla_{\mathbf{r}} v_{\text{ray}} = \begin{bmatrix} +3.3593811\cdot 10^{-1} & -2.1127882\cdot 10^{-1} & +6.7079628\cdot 10^{-2} \\ -8.5407762\cdot 10^{-2} & -2.0423345\cdot 10^{-1} & +1.8446798\cdot 10^{-1} \\ -4.5565394\cdot 10^{-2} & +8.6509905\cdot 10^{-2} & -5.4295956\cdot 10^{-2} \end{bmatrix} \quad \text{s}^{-1} \quad , \qquad (F29)$$

and,

$$\nabla_{\mathbf{r}}\nabla_{\mathbf{x}} v_{\text{ray}} = \begin{bmatrix} +3.3593811\cdot 10^{-1} & -8.5407762\cdot 10^{-2} & -4.5565394\cdot 10^{-2} \\ -2.1127882\cdot 10^{-1} & -2.0423345\cdot 10^{-1} & +8.6509905\cdot 10^{-2} \\ +6.7079628\cdot 10^{-2} & +1.8446798\cdot 10^{-1} & -5.4295956\cdot 10^{-2} \end{bmatrix} \quad \text{s}^{-1} \quad . \qquad (F30)$$

The computational formulae for the gradients and Hessians of the slowness vector and the ray velocity magnitude are provided in Appendix E of Part I.

## APPENEIX G.

## LAGRANGIAN/HAMILTONIAN TEST FOR ELLIPSOIDAL ANISOTROPY

In this appendix we perform analytically the test which was done in Appendices E and F numerically, now for an ellipsoidal anisotropy. The ellipsoidal anisotropy is defined by three parameters $A_v, B_v, C_v$ representing the axial velocities in the crystal frame for a given wave mode. In this analysis, we assume that the global frame coincides with the crystal frame. The Legendre transform for this medium has been analyzed by Červený (2002b), page 229.

The slowness surface of the ellipsoidal anisotropy is given by (Červený, 2002b), equation 95,

$$A_v^2(\mathbf{x}) p_1^2 + B_v^2(\mathbf{x}) p_2^2 + C_v^2(\mathbf{x}) p_3^2 = 1 \qquad , \qquad (G1)$$

and the ellipsoidal Hamiltonian reads,

$$H^e(\mathbf{x},\mathbf{p}) = \frac{1}{2}\left[ A_v^2(\mathbf{x})p_1^2 + B_v^2(\mathbf{x})p_2^2 + C_v^2(\mathbf{x})p_3^2 - 1 \right] \quad . \tag{G2}$$

The flow variable of the ellipsoidal Hamiltonian $H^e(\mathbf{x},\mathbf{p})$ is the traveltime. In the original paper by Červený (2002b), equation 94, 1 is not subtracted as in our equation G2, therefore, the Hamiltonian of the cited for ellipsoidal media is $1/2$ (in the same way as the eigenvalue Hamiltonian $H^\lambda$ for general anisotropic media), while our Hamiltonian $H^e$ vanishes (in the same way as Hamiltonians $H^{\bar{\tau}}, H^s, H^\tau, H^\sigma$). The ellipsoidal anisotropy can be viewed as a particular case of orthorhombic symmetry. If two of the axial velocities coincide, it becomes a particular case of transverse isotropy (TI), referred also as elliptic polar anisotropy. In the case where all axial velocities are equal, the medium is isotropic. We assume here that parameters $A_v, B_v, C_v$ are all different.

The ellipsoidal medium is a particular case of an (acoustic) orthorhombic medium with the following axial velocities and constraints,

$$\begin{aligned} C_{11} &= A_v^2 & C_{12} &= \sqrt{C_{11}C_{22}} & C_{44} &= 0 \\ C_{22} &= B_v^2 & C_{13} &= \sqrt{C_{11}C_{33}} & C_{55} &= 0 \\ C_{33} &= C_v^2 & C_{23} &= \sqrt{C_{22}C_{33}} & C_{66} &= 0 \end{aligned} \quad . \tag{G3}$$

With these elastic parameters, the reference Hamiltonian $H^{\bar{\tau}}$ of equation 11 of Part I and its corresponding scaled time become,

$$H^{\bar{\tau}}(\mathbf{x},\mathbf{p}) = 2H^e(\mathbf{x},\mathbf{p}) \quad , \quad \bar{\tau} = \tau/2 \quad . \tag{G4}$$

We start from the transform that can be considered an equivalent to the Legendre transform (or its replacement). Given the ray velocity direction $\mathbf{r}$, find the slowness $\mathbf{p}$. For this we first establish the slowness gradient of the Hamiltonian,

$$H^e_{\mathbf{p}} = \left[ A_v^2(\mathbf{x}) p_1 \quad B_v^2(\mathbf{x}) p_2 \quad C_v^2(\mathbf{x}) p_3 \right] \quad . \tag{G5}$$

Note that for this case,

$$\mathbf{p} \cdot H^e_{\mathbf{p}} = 1 \quad \rightarrow \quad \mathbf{v}_{\text{ray}} = \frac{H^e_{\mathbf{p}}}{\mathbf{p} \cdot H^e_{\mathbf{p}}} = H^e_{\mathbf{p}} \quad . \tag{G6}$$

Note also that since $\mathbf{v}_{\text{ray}} = H^e_{\mathbf{p}}$, then,

$$\frac{d\mathbf{x}}{ds} = \frac{H^e_{\mathbf{p}}}{\sqrt{H^e_{\mathbf{p}} \cdot H^e_{\mathbf{p}}}} = \frac{\mathbf{v}_{\text{ray}}}{v_{\text{ray}}} = \mathbf{r} \quad , \tag{G7}$$

i.e., the Lagrangian and Hamiltonian definitions of the ray velocity direction are identical.

Next, we solve equation set 18 of Part I for the unknown slowness components, applying the ellipsoidal Hamiltonian, $H^e(\mathbf{x}, \mathbf{p})$, from equation G2. The solution reads,

$$\mathbf{p} = \frac{1}{\sqrt{B_v^2(\mathbf{x}) C_v^2(\mathbf{x}) r_1^2 + A_v^2(\mathbf{x}) C_v^2(\mathbf{x}) r_2^2 + A_v^2(\mathbf{x}) B_v^2(\mathbf{x}) r_3^2}} \begin{bmatrix} \dfrac{B_v(\mathbf{x}) C_v(\mathbf{x})}{A_v(\mathbf{x})} r_1 \\ \dfrac{A_v(\mathbf{x}) C_v(\mathbf{x})}{B_v(\mathbf{x})} r_2 \\ \dfrac{A_v(\mathbf{x}) B_v(\mathbf{x})}{C_v(\mathbf{x})} r_3 \end{bmatrix} \tag{G8}$$

Thus, given the ray direction $\mathbf{r}$, one can find the slowness $\mathbf{p}$. The inverse relationship exists as well: given the slowness, one can find the ray direction,

$$\mathbf{r} = \frac{1}{\sqrt{A_v^4(\mathbf{x})p_1^2 + B_v^4(\mathbf{x})p_2^2 + C_v^4(\mathbf{x})p_3^2}} \begin{bmatrix} A_v^2(\mathbf{x})p_1 \\ B_v^2(\mathbf{x})p_2 \\ C_v^2(\mathbf{x})p_3 \end{bmatrix} . \tag{G9}$$

Introduction of equation G8 into G5 and G6 yields the magnitude of the ray velocity in terms of its direction,

$$v_{\text{ray}}(\mathbf{x},\mathbf{r}) = \frac{A_v(\mathbf{x})B_v(\mathbf{x})C_v(\mathbf{x})}{\sqrt{B_v^2(\mathbf{x})C_v^2(\mathbf{x})r_1^2 + A_v^2(\mathbf{x})C_v^2(\mathbf{x})r_2^2 + A_v^2(\mathbf{x})B_v^2(\mathbf{x})r_3^2}} . \tag{G10}$$

We emphasize that equation G10 is not an evidence that the ray velocity magnitude reciprocal is first-degree homogeneous wrt the tangent vector $\mathbf{r}$, because this vector is necessarily normalized. Should we use a flow parameter $\zeta$, other than the arclength and leading to the non-normalized tangent vector, $\dot{\mathbf{x}}_\zeta = d\mathbf{x}/d\zeta = k(\zeta)\mathbf{r}$, equation G10 transforms into,

$$\begin{aligned} v_{\text{ray}}(\mathbf{x},\dot{\mathbf{x}}_\zeta) &= \frac{A_v(\mathbf{x})B_v(\mathbf{x})C_v(\mathbf{x})}{\sqrt{B_v^2(\mathbf{x})C_v^2(\mathbf{x})\frac{\dot{x}_{\zeta,1}^2}{\dot{\mathbf{x}}_\zeta\cdot\dot{\mathbf{x}}_\zeta} + A_v^2(\mathbf{x})C_v^2(\mathbf{x})\frac{\dot{x}_{\zeta,2}^2}{\dot{\mathbf{x}}_\zeta\cdot\dot{\mathbf{x}}_\zeta} + A_v^2(\mathbf{x})B_v^2(\mathbf{x})\frac{\dot{x}_{\zeta,3}^2}{\dot{\mathbf{x}}_\zeta\cdot\dot{\mathbf{x}}_\zeta}}} \\ &= \frac{A_v(\mathbf{x})B_v(\mathbf{x})C_v(\mathbf{x})\sqrt{\dot{\mathbf{x}}_\zeta\cdot\dot{\mathbf{x}}_\zeta}}{\sqrt{B_v^2(\mathbf{x})C_v^2(\mathbf{x})\dot{x}_{\zeta,1}^2 + A_v^2(\mathbf{x})C_v^2(\mathbf{x})\dot{x}_{\zeta,2}^2 + A_v^2(\mathbf{x})B_v^2(\mathbf{x})\dot{x}_{\zeta,3}^2}} , \end{aligned} \tag{G11}$$

clearly indicating that both, the ray velocity magnitude and its reciprocal are zero-degree homogeneous functions wrt vector $\dot{\mathbf{x}}_\zeta$.

The ray velocity magnitude in terms of slowness components reads,

$$v_{\text{ray}}(\mathbf{x},\mathbf{p}) = \sqrt{A_v^4(\mathbf{x})p_1^2 + B_v^4(\mathbf{x})p_2^2 + C_v^4(\mathbf{x})p_3^2} \qquad . \qquad (G12)$$

It follows from equation G10 that,

$$\frac{r_1^2}{A_v^2(\mathbf{x})} + \frac{r_2^2}{B_v^2(\mathbf{x})} + \frac{r_3^2}{C_v^2(\mathbf{x})} = \frac{1}{v_{\text{ray}}^2(\mathbf{x},\mathbf{r})} \qquad . \qquad (G13)$$

This equation was obtained by Červený (2002b), equation 103, in a different way and slightly different form.

Equation G8 yields the phase velocity vs. the ray direction,

$$v_{\text{phs}}(\mathbf{x},\mathbf{r}) = A_v(\mathbf{x})B_v(\mathbf{x})C_v(\mathbf{x}) \frac{\sqrt{B_v^2(\mathbf{x})C_v^2(\mathbf{x})r_1^2 + A_v^2(\mathbf{x})C_v^2(\mathbf{x})r_2^2 + A_v^2(\mathbf{x})B_v^2(\mathbf{x})r_3^2}}{\sqrt{B_v^4(\mathbf{x})C_v^4(\mathbf{x})r_1^2 + A_v^4(\mathbf{x})C_v^4(\mathbf{x})r_2^2 + A_v^4(\mathbf{x})B_v^4(\mathbf{x})r_3^2}} \qquad , \qquad (G14)$$

and the ratio,

$$\frac{v_{\text{phs}}(\mathbf{x},\mathbf{r})}{v_{\text{ray}}(\mathbf{x},\mathbf{r})} = \frac{B_v^2(\mathbf{x})C_v^2(\mathbf{x})r_1^2 + A_v^2(\mathbf{x})C_v^2(\mathbf{x})r_2^2 + A_v^2(\mathbf{x})B_v^2(\mathbf{x})r_3^2}{\sqrt{B_v^4(\mathbf{x})C_v^4(\mathbf{x})r_1^2 + A_v^4(\mathbf{x})C_v^4(\mathbf{x})r_2^2 + A_v^4(\mathbf{x})B_v^4(\mathbf{x})r_3^2}} \le 1 \qquad . \qquad (G15)$$

With equation G10, we compute the spatial gradient of the ray velocity magnitude,

$$\nabla_{\mathbf{x}} v_{\text{ray}} = -\frac{\nabla_{\mathbf{x}} A_v(\mathbf{x}) B_v^3(\mathbf{x}) C_v^3(\mathbf{x}) r_1^2 + A_v^3(\mathbf{x}) \nabla_{\mathbf{x}} B_v(\mathbf{x}) C_v^3(\mathbf{x}) r_2^2 + A_v^3(\mathbf{x}) B_v^3(\mathbf{x}) \nabla_{\mathbf{x}} C_v(\mathbf{x})}{\left[ B_v^2(\mathbf{x}) C_v^2(\mathbf{x}) r_1^2 + A_v^2(\mathbf{x}) C_v^2(\mathbf{x}) r_2^2 + A_v^2(\mathbf{x}) B_v^2(\mathbf{x}) r_3^2 \right]^{3/2}} \qquad . (G16)$$

We compute the derivative of the slowness vector wrt the arclength that follows from the arclength-related Lagrangian $L(s)$,

$$\left(\frac{d\mathbf{p}}{ds}\right)_L = -\frac{\nabla_{\mathbf{x}} v_{\text{ray}}}{v_{\text{ray}}^2} =$$
$$-\frac{\nabla_{\mathbf{x}} A_v(\mathbf{x}) B_v^3(\mathbf{x}) C_v^3(\mathbf{x}) r_1^2 + A_v^3(\mathbf{x}) \nabla_{\mathbf{x}} B_v(\mathbf{x}) C_v^3(\mathbf{x}) r_2^2 + A_v^3(\mathbf{x}) B_v^3(\mathbf{x}) \nabla_{\mathbf{x}} C_v(\mathbf{x})}{A_v^2(\mathbf{x}) B_v^2(\mathbf{x}) C_v^2(\mathbf{x}) \sqrt{B_v^2(\mathbf{x}) C_v^2(\mathbf{x}) r_1^2 + A_v^2(\mathbf{x}) C_v^2(\mathbf{x}) r_2^2 + A_v^2(\mathbf{x}) B_v^2(\mathbf{x}) r_3^2}} \quad . \tag{G17}$$

Next, we compute the same derivative from the scaled arclength-related Hamiltonian $H(\mathbf{x},\mathbf{p})$,

$$\left(\frac{d\mathbf{p}}{ds}\right)_H = -H_{\mathbf{x}} = -\frac{H_{\mathbf{x}}^e(\mathbf{x},\mathbf{p})}{\sqrt{H_{\mathbf{p}}^e(\mathbf{x},\mathbf{p}) \cdot H_{\mathbf{p}}^e(\mathbf{x},\mathbf{p})}} =$$
$$-\frac{A_v(\mathbf{x}) \nabla_{\mathbf{x}} A_v(\mathbf{x}) p_1^2 + B_v(\mathbf{x}) \nabla_{\mathbf{x}} B_v(\mathbf{x}) C_v^3(\mathbf{x}) p_2^2 + C_v(\mathbf{x}) \nabla_{\mathbf{x}} C_v(\mathbf{x}) p_3^2}{\sqrt{A_v^4(\mathbf{x}) p_1^2 + B_v^4(\mathbf{x}) p_2^2 + C_v^4(\mathbf{x}) p_3^2}} \quad . \tag{G18}$$

Next, we compute the same derivative, applying the eigenvalue Hamiltonian (equation 30) suggested by Červený (2000, 2002a, 2002b) and compute the polarization $\mathbf{g}(\mathbf{x},\mathbf{p})$ from the Christoffel matrix,

$$\mathbf{g}(\mathbf{x},\mathbf{p}) = \frac{\left[A_w(\mathbf{x}) p_1 \quad B_w(\mathbf{x}) p_2 \quad C_w(\mathbf{x}) p_3\right]}{\sqrt{A_v^2(\mathbf{x}) p_1^2 + B_v^2(\mathbf{x}) p_2^2 + C_v^2(\mathbf{x}) p_3^2}} \quad . \tag{G19}$$

This polarization vector corresponds to the eigenvalue $\lambda = 1$. Although the denominator on the right-hand side of equation G19 is identically 1, it is worth keeping it to obtain a simpler expression for the eigenvalue Hamiltonian $H^\lambda$ in equation 30. The polarization can be also presented in terms of the ray direction components,

$$\mathbf{g}(\mathbf{x},\mathbf{r}) = \frac{\left[B_w(\mathbf{x}) C_w(\mathbf{x}) r_1 \quad A_w(\mathbf{x}) C_w(\mathbf{x}) r_2 \quad A_w(\mathbf{x}) B_w(\mathbf{x}) p_3\right]}{\sqrt{B_v^2(\mathbf{x}) C_v^2(\mathbf{x}) r_1^2 + A_v^2(\mathbf{x}) C_v^2(\mathbf{x}) p_2^2 + A_v^2(\mathbf{x}) B_v^2(\mathbf{x}) p_3^2}} \quad . \tag{G20}$$

We then make use of equation B15 and obtain the same right-hand side as in equation G18. As expected, both the arclength-related and the eigenvalue Hamiltonians lead to identical results,

$$\left(\frac{d\mathbf{p}}{ds}\right)_H = \left(\frac{d\mathbf{p}}{ds}\right)_{H^\lambda} . \tag{G21}$$

Finally, we introduce our kind of "Legendre transform", $\mathbf{p} = \mathbf{p}(\mathbf{x}, \mathbf{r})$, equation G8, into G18, and we obtain the derivative $d\mathbf{p}/ds$ from the Hamiltonian, but in terms of the ray velocity direction $\mathbf{r}$, rather than in terms of the slowness $\mathbf{p}$. This leads us to equation G17. Thus, we demonstrated analytically that for ellipsoidal media,

$$\left(\frac{d\mathbf{p}}{ds}\right)_L = \left(\frac{d\mathbf{p}}{ds}\right)_H = \left(\frac{d\mathbf{p}}{ds}\right)_{H^\lambda} , \tag{G22}$$

where $H(\mathbf{x}, \mathbf{p})$ is the proposed arclength-related Hamiltonian that for ellipsoidal media simplifies to $H = H^e / \sqrt{H^e_\mathbf{p} \cdot H^e_\mathbf{p}}$, and $H^\lambda(\mathbf{x}, \mathbf{p})$ is the eigenvalue Hamiltonian suggested by Červený (equation 30).

We emphasize that identity G22 holds for any anisotropic media (as we demonstrated numerically for the most general anisotropic case with all stiffness components varying in 3D space). The derivative of the slowness components wrt the arclength computed with the use of the Lagrangian and Hamiltonian, are identical. The same is true for the derivative of the ray location wrt the arclength, i.e., both Lagrangian and Hamiltonian yield the same ray velocity direction $d\mathbf{x}/ds = \mathbf{r}$.

Next, we find the directional gradient of the ray velocity. The non-normalized directional gradient reads,

$$\frac{\partial v_{\text{ray}}(\mathbf{x},\mathbf{r})}{\partial \mathbf{r}} = -\frac{A_v(\mathbf{x})B_v(\mathbf{x})C_v(\mathbf{x})}{\left[B_v^2(\mathbf{x})C_v^2(\mathbf{x})r_1^2 + A_v^2(\mathbf{x})C_v^2(\mathbf{x})r_2^2 + A_v^2(\mathbf{x})B_v^2(\mathbf{x})r_3^2\right]^{3/2}} \begin{bmatrix} B_v^2(\mathbf{x})C_v^2(\mathbf{x})r_1 \\ A_v^2(\mathbf{x})C_v^2(\mathbf{x})r_2 \\ A_v^2(\mathbf{x})B_v^2(\mathbf{x})r_3 \end{bmatrix}.$$

(G23)

The normalized directional gradient reads,

$$\nabla_{\mathbf{r}} v_{\text{ray}}(\mathbf{x},\mathbf{r}) = (\mathbf{I} - \mathbf{r} \otimes \mathbf{r})\frac{\partial v_{\text{ray}}(\mathbf{x},\mathbf{r})}{\partial \mathbf{r}} =$$

$$= \frac{A_v(\mathbf{x})B_v(\mathbf{x})C_v(\mathbf{x})}{\left[B_v^2(\mathbf{x})C_v^2(\mathbf{x})r_1^2 + A_v^2(\mathbf{x})C_v^2(\mathbf{x})r_2^2 + A_v^2(\mathbf{x})B_v^2(\mathbf{x})r_3^2\right]^{3/2}} \quad (G24)$$

$$\cdot \begin{Bmatrix} C_v^2(\mathbf{x})\left[A_v^2(\mathbf{x}) - B_v^2(\mathbf{x})\right]r_1 r_2^2 + B_v^2(\mathbf{x})\left[A_v^2(\mathbf{x}) - C_v^2(\mathbf{x})\right]r_1 r_3^2 \\ C_v^2(\mathbf{x})\left[B_v^2(\mathbf{x}) - A_v^2(\mathbf{x})\right]r_1^2 r_2 + A_v^2(\mathbf{x})\left[B_v^2(\mathbf{x}) - C_v^2(\mathbf{x})\right]r_2 r_3^2 \\ B_v^2(\mathbf{x})\left[C_v^2(\mathbf{x}) - A_v^2(\mathbf{x})\right]r_1^2 r_3 + A_v^2(\mathbf{x})\left[C_v^2(\mathbf{x}) - B_v^2(\mathbf{x})\right]r_2^2 r_3 \end{Bmatrix}.$$

One can see that for an isotropic medium, $A_v(\mathbf{x}) = B_v(\mathbf{x}) = C_v(\mathbf{x})$, all square brackets in the numerator vanish, and the normalized directional gradient of the ray velocity in equation G24 is zero. This is not so for the non-normalized directional gradient in equation G23, which does not vanish for an isotropic medium. One can also see from equation G24 that the normalized directional gradient of the ray velocity is normal to the ray velocity direction, $\nabla_{\mathbf{r}} v_{\text{ray}}(\mathbf{x},\mathbf{r}) \cdot \mathbf{r} = 0$.

Next, we compute the generalized momentum using equations A2 of Part I, G10 and G24, and we make sure that we obtain the slowness vector as in equation G8.

Compute tensor $\mathbf{F}$ that relates the ray velocity vector to the slowness vector,

$$\mathbf{F} = \mathbf{g}\tilde{\mathbf{C}}\mathbf{g} = \begin{bmatrix} A_w^4(\mathbf{x})p_1^2 & A_w^2(\mathbf{x})B_w^2(\mathbf{x})p_1 p_2 & A_w^2(\mathbf{x})C_w^2(\mathbf{x})p_1 p_3 \\ A_w^2(\mathbf{x})B_w^2(\mathbf{x})p_1 p_2 & B_w^4(\mathbf{x})p_2^2 & B_w^2(\mathbf{x})C_w^2(\mathbf{x})p_2 p_3 \\ A_w^2(\mathbf{x})C_w^2(\mathbf{x})p_1 p_3 & B_w^2(\mathbf{x})C_w^2(\mathbf{x})p_2 p_3 & C_w^4(\mathbf{x})p_3^2 \end{bmatrix} \qquad (G25)$$

$$= H_\mathbf{p}^e \otimes H_\mathbf{p}^e = \mathbf{v}_{\text{ray}} \otimes \mathbf{v}_{\text{ray}}$$

Tensor (matrix) $\mathbf{F}$ has an eigenvalue $v_{\text{ray}}^2$, and the corresponding eigenvector is the ray direction $\mathbf{r}$. The two other eigenvalues are zero, with their eigenvectors normal to the ray and to each other. Multiplication of this matrix by the slowness vector yields the ray velocity vector,

$$\mathbf{F}\mathbf{p} = \begin{bmatrix} A_w^2(\mathbf{x})p_1 & B_w^2(\mathbf{x})p_2 & C_w^2(\mathbf{x})p_3 \end{bmatrix}^T = H_\mathbf{p}^e = \mathbf{v}_{\text{ray}} \qquad (G26)$$

The inverse of the matrix in equation G25 does not exist. Thus, with this tensor, we can obtain the ray velocity from the slowness, but not vice versa. The reason is that the Hamiltonian of the ellipsoidal anisotropy in equation G2 assumes implicitly the acoustic approximation. Indeed, it describes only one wave mode which can be considered, for example, compressional.

However, the two-way relationship becomes possible with the use of the Finsler metric $\mathbf{G}$ related to the Hessian $H_{\mathbf{pp}}^\lambda$ suggested by Červený (2002a, 2002b) that results in a diagonal matrix, whose determinant does not vanish,

$$\mathbf{G}^{-1} = H_{\mathbf{pp}}^e = \begin{bmatrix} A_w^2(\mathbf{x}) & 0 & 0 \\ 0 & B_w^2(\mathbf{x}) & 0 \\ 0 & 0 & C_w^2(\mathbf{x}) \end{bmatrix} \qquad (G27)$$

In this case, $\mathbf{p} \cdot H_{\mathbf{p}}^e = 1$ and $\mathbf{p} H_{\mathbf{pp}}^e \mathbf{p} = 1$, therefore, the use of equation C8 leads to the same result.

Consider the difference matrix $\mathbf{A}$,

$$\mathbf{A} = \mathbf{F} - \mathbf{G}^{-1} = \mathbf{g}\tilde{\mathbf{C}}\mathbf{g} - \mathbf{G}^{-1} \quad . \tag{G28}$$

Each row of matrix $\mathbf{A}$ is normal to the slowness direction, so that,

$$\mathbf{A}\mathbf{p} = 0 \quad \rightarrow \quad \mathbf{G}^{-1}\mathbf{p} = \mathbf{F}\mathbf{p} = \mathbf{v}_{\text{ray}} \;, \quad \mathbf{G}^{-1} \neq \mathbf{F} \quad . \tag{G29}$$

The inverse matrix $\mathbf{F}^{-1}$ in this case does not exist, while the Finsler metric $\mathbf{G}$ exists and performs the inverse transform from the ray velocity vector to the slowness vector, $\mathbf{G}\mathbf{v}_{\text{ray}} = \mathbf{p}$.

## APPENDIX H. SOLVING MOMENTUM EQUATION FOR RAY DIRECTION

In this appendix, we solve the generalized momentum equation,

$$L^\zeta_{\dot{\mathbf{x}}_\xi}\left(\mathbf{x}, \dot{\mathbf{x}}_\xi\right) = \mathbf{p} \quad , \tag{H1}$$

for the unknown position derivative, $\dot{\mathbf{x}}_\xi \equiv d\mathbf{x}/d\zeta$, where $\zeta$ is an arbitrary flow parameter (e.g., time, arclength or sigma), given a general anisotropic elastic tensor $\tilde{\mathbf{C}}(\mathbf{x})$ at a specified location $\mathbf{x}$, and the slowness vector $\mathbf{p}$ at that location. In particular, we are primarily interested and refer to the arclength-related case, $\zeta \to s$, where $\dot{\mathbf{x}}_\xi$ is the ray velocity direction $\mathbf{r} = \dot{\mathbf{x}} = \dot{\mathbf{x}}_s \equiv d\mathbf{x}/ds$. (Note that the inverse problem for obtaining the slowness vector given the ray velocity direction is discussed in Appendix D of Part I.) The ray direction can be solved by applying either the proposed arclength-related Hamiltonian, $H(\mathbf{x}, \mathbf{p})$, or, alternatively its accompanying (connected through the Legendre transform) Lagrangian, $L(\mathbf{x},\mathbf{r})$. The computation of the ray direction using the Hamiltonian-based kinematic equation, $\dot{\mathbf{x}} = \mathbf{r} = H_{\mathbf{p}}(\mathbf{x},\mathbf{p})$, is straightforward, and this is our actual solution method in this study. On the other hand, the Lagrangian-based approach for obtaining the ray direction requires solving the momentum equation, $L_{\mathbf{r}}(\mathbf{x},\mathbf{r}) = \mathbf{p}$, which is a nonlinear (inversion) equation and can only be solved numerically (iteratively). The nonlinearity is due to the dependency of the directional gradient of the proposed Lagrangian, $L_{\mathbf{r}}$, on the corresponding directional gradient of the ray velocity magnitude, $\nabla_{\mathbf{r}} v_{\text{ray}}$, which in turn, depends nonlinearly on the ray direction vector $\mathbf{r}$. Combining equations A14 and A15 of Part I, we obtain,

$$L_{\mathbf{r}} = \mathbf{p} \quad \text{or} \quad (\mathbf{p}\cdot\mathbf{r})\mathbf{r} + \mathbf{r}\times\mathbf{p}\times\mathbf{r} = \mathbf{p} \tag{H2}$$

Equation H2 shows that the slowness vector can be decomposed in two components: lengthwise, along the ray, and transverse, in the plane normal to the ray. This decomposition can be applied when both, the slowness $L_{\mathbf{r}} = \mathbf{p}$ and the ray direction $\mathbf{r}$ are true known *compatible* ray vectors. The right relationship of equation H2, however, does not allow to establish the ray direction, given the slowness. One can choose any direction and decompose the slowness vector into its lengthwise and transverse components. The relationship is correct, but it does not contain the medium properties and hence cannot be considered the momentum equation; it is just an identity. The reason is that in the derivation of the directional gradient of the ray velocity, $\nabla_{\mathbf{r}} v_{\text{ray}}$ and its Hessian, $\nabla_{\mathbf{r}}\nabla_{\mathbf{r}} v_{\text{ray}}$, that we need for the numerical solution of the momentum equation, we assumed the abovementioned compatibility of the ray direction and the slowness vectors. It means that the ray direction $\mathbf{r}$ is collinear with the slowness gradient of (any) Hamiltonian, $H_{\mathbf{p}}^{\zeta}$.

Computing the directional gradient $L_{\mathbf{r}}$ and Hessian $L_{\mathbf{rr}}$ of the arclength-related Lagrangian

In order to include the medium properties, we need first to review the derivation of the directional gradient and Hessian of the ray velocity, $\nabla_{\mathbf{r}} v_{\text{ray}}$ and $\nabla_{\mathbf{r}} \nabla_{\mathbf{r}} v_{\text{ray}}$, (Ravve and Koren, 2019), required to compute the corresponding gradient and Hessian of the Lagrangian, $L_{\mathbf{r}}$ and $L_{\mathbf{rr}}$. The key objects in this derivation are the second- and third- order tensors, $\mathbf{p_r} = \partial \mathbf{p} / \partial \mathbf{r}$ and $\mathbf{p_{rr}} = \partial^2 \mathbf{p} / \partial \mathbf{r}^2$, needed for the directional gradient and Hessian of the ray velocity, respectively. Their derivation is explained in details in the cited paper. The derivation is based on the symmetry of the second-order tensor $\mathbf{p_r}$ and its orthogonality to the ray direction $\mathbf{r}$, which in turn, lead to the super-symmetry (insensitivity to any index permutations) of the third-order tensor $\mathbf{p_{rr}}$. When the ray direction is slowness-compatible, then $L_{\mathbf{r}} = \mathbf{p}$ and $L_{\mathbf{rr}} = \mathbf{p_r}$, otherwise these identities do not hold, and the mentioned properties of the slowness gradient and Hessian tensors $\mathbf{p_r}$ and $\mathbf{p_{rr}}$ do not hold either. In this appendix, we apply "more general" relationships that approximate the directional gradient and Hessian of the ray velocity for the "incompatible" formulation. We will demonstrate, which terms simplify as the ray direction converges to that of the true ray during the iterative solution of the momentum equation.

Hence, we suggest the following workflow to compute $L_{\mathbf{r}}$ and $L_{\mathbf{rr}}$.

1. Given the medium properties, the slowness vector and the approximate ray velocity direction, we design the functions, $H_{\mathbf{p}}^{\bar{\tau}} \times \mathbf{r}_{\text{o}}$ and $H^{\bar{\tau}}(\mathbf{x}, \mathbf{p})$. Any Hamiltonian can be

Page 65 of 87

applied; we suggest the simplest one which is the reference Hamiltonian $H^{\bar{\tau}}$. Since the three slowness components are compatible with each other, the Hamiltonian vanishes; however, its slowness gradient $H_{\mathbf{p}}^{\bar{\tau}}$ is collinear with the actual ray direction $\mathbf{r}$ rather than with the approximation $\mathbf{r}_o$, such that $H_{\mathbf{p}}^{\bar{\tau}} \times \mathbf{r}_o \neq 0$. Since the three Cartesian components of the cross product are dependent, we apply two of them and for the third equation we use the vanishing gradient. For a fixed location $\mathbf{x}$, this set constitutes a vector-form function $\mathbf{f}(\mathbf{p},\mathbf{r})$, with three components, $f_k$ (Ravve and Koren, 2019, equation 16),

$$\mathbf{f}(\mathbf{p},\mathbf{r}) = \begin{bmatrix} \left(H_{\mathbf{p}}^{\bar{\tau}} \times \mathbf{r}_o\right)_1 \\ \left(H_{\mathbf{p}}^{\bar{\tau}} \times \mathbf{r}_o\right)_2 \\ H^{\bar{\tau}}(\mathbf{x},\mathbf{p}) \end{bmatrix} \quad . \tag{H3}$$

When solving the momentum equation, we do not solve explicitly for $\mathbf{f} = 0$; however, this vector will vanish on the completion of the iterative procedure, when the true ray direction is found. Note also that the third component of $\mathbf{f}$ (the Hamiltonian) vanishes at all iterations as we update only the ray direction and do not change the slowness.

2. Compute the second-order gradients and the third-order Hessians of vector $\mathbf{f}(\mathbf{p},\mathbf{r})$,

$$\mathbf{f}_{\mathbf{p}}, \mathbf{f}_{\mathbf{r}}, \mathbf{f}_{\mathbf{pp}}, \mathbf{f}_{\mathbf{rr}}, \mathbf{f}_{\mathbf{pr}}, \mathbf{f}_{\mathbf{rp}} \quad . \tag{H4}$$

In the gradients, the first index is related to the component of vector $\mathbf{f}$, and the second index – to the derivative component. In the Hessians, the first index is related to the component of $\mathbf{f}$, and the two other indices are related to the derivatives.

3. Compute the second-order tensor of the slowness gradient wrt the ray direction,

$$\mathbf{p_r} = -\mathbf{f_p}^{-1}\mathbf{f_r} \tag{H5}$$

The right-hand side of this equation is a product of two matrices; each of them and the resulting matrix have dimensions $3\times 3$. Equation H5 follow from equation 19 of the cited paper, we arranged now it in the tensor form.

4. Compute the third-order tensor of the slowness Hessian wrt the ray direction,

$$\mathbf{p_{rr}} = -\mathbf{f_p}^{-1}\left[\left(\mathbf{p_r}^T \mathbf{f_{pp}}^{T\{2,1,3\}} \mathbf{p_r}\right)^{T\{2,1,3\}} + \left(\mathbf{p_r}^T \mathbf{f_{pr}}^{T\{2,1,3\}}\right)^{T\{2,1,3\}} + \mathbf{f_{rp}}\mathbf{p_r} + \mathbf{f_{rr}}\right] . \tag{H6}$$

Equation H6, arranged here in a tensor form, follows from equations 23 and 26 of the cited paper. Transposed $T\{2,1,3\}$ for the third-order tensor means that indices 1 and 2 are swapped. This is needed to get access to its second, "internal", index, when multiplying with the other tensor from the left. On the completion of the operation, the indices of the third-order resulting tensor are returned to their original positions, therefore we see the transpose operator twice: inside and outside the round brackets.

The last term in the square brackets, $\mathbf{f_{rr}}$, vanishes because the first two components of vector $\mathbf{f}$ in equation H3 depend linearly on the ray direction, and the third component is independent of the ray direction.

5. Compute the approximation for the ray velocity magnitude and its non-normalized directional gradient,

$$v_{ray} = \frac{1}{\mathbf{p} \cdot \mathbf{r}_o} \quad , \quad \frac{\partial v_{ray}}{\partial \mathbf{r}} = -v_{ray}^2 \left( \mathbf{p} + \mathbf{r}_o \, \mathbf{p_r} \right) \quad . \tag{H7}$$

The second equation of set H7 is similar to equation 21 in the cited paper, but here we take into account that for "incompatible case", the directional gradient of the slowness, $\mathbf{p_r}$ in not a symmetric tensor for the "incompatible" case.

6. Compute the approximation for the non-normalized Hessian of the ray velocity magnitude,

$$\frac{\partial^2 v_{ray}}{\partial \mathbf{r}^2} = \frac{2}{v_{ray}} \frac{\partial v_{ray}}{\partial \mathbf{r}} \otimes \frac{\partial v_{ray}}{\partial \mathbf{r}} - v_{ray}^2 \left( \mathbf{p_r} + \mathbf{p_r}^T + \mathbf{r}_o \, \mathbf{p_{rr}} \right) \quad . \tag{H8}$$

Equation H8 is similar to equation 27 of the cited paper, but here we take into account that for "incompatible case", the slowness gradient, $\mathbf{p_r}$ is not a symmetric tensor, and the slowness Hessian, $\mathbf{p_{rr}}$, is not super-symmetric (in other words, only two last indices of $\mathbf{p_{rr}}$ related to the derivative components can be swapped, but not the first index, related to the slowness components).

7. Normalize the directional gradient of the ray velocity,

$$\begin{aligned} \nabla_{\mathbf{r}} v_{ray} &= \mathbf{T} \frac{\partial v_{ray}}{\partial \mathbf{r}} \quad , \\ \nabla_{\mathbf{r}} \nabla_{\mathbf{r}} v_{ray} &= \mathbf{E} \frac{\partial v_{ray}}{\partial \mathbf{r}} + \mathbf{T} \frac{\partial^2 v_{ray}}{\partial \mathbf{r}^2} \mathbf{T} \quad , \end{aligned} \tag{H9}$$

where tensors **T** and **E** are defined in equations E3 and E15 of Part I, respectively.

8. Next we compute the directional gradient and Hessian of the arclength-related Lagrangian $L$ (in fact, their approximations, because $\mathbf{r}_o$ is introduced instead of the true ray direction $\mathbf{r}$ to equations H10 and H11),

$$L_{\mathbf{r}} = \frac{\mathbf{r}}{v_{\text{ray}}} - \frac{\nabla_{\mathbf{r}} v_{\text{ray}}}{v_{\text{ray}}^2} \quad , \tag{H10}$$

$$L_{\mathbf{rr}} = \frac{\mathbf{I} - \mathbf{r} \otimes \mathbf{r}}{v_{\text{ray}}} - \frac{\mathbf{r} \otimes \nabla_{\mathbf{r}} v_{\text{ray}} + \nabla_{\mathbf{r}} v_{\text{ray}} \otimes \mathbf{r}}{v_{\text{ray}}^2} + 2 \frac{\nabla_{\mathbf{r}} v_{\text{ray}} \otimes \nabla_{\mathbf{r}} v_{\text{ray}}}{v_{\text{ray}}^3} - \frac{\nabla_{\mathbf{r}} \nabla_{\mathbf{r}} v_{\text{ray}}}{v_{\text{ray}}^2} \quad . \tag{H11}$$

9. Combining equations H7, H9 and H10, we obtain the momentum equation for the "incompatible" case,

$$L_{\mathbf{r}} - \mathbf{p} = \mathbf{T}\left(\mathbf{p}_{\mathbf{r}}^T \mathbf{r}_o\right) = \left(\mathbf{I} - \mathbf{r} \otimes \mathbf{r}\right)\left(\mathbf{p}_{\mathbf{r}}^T \mathbf{r}_o\right) = 0 \quad . \tag{H12}$$

As we proceed with the iterations and the approximation $\mathbf{r}_o$ approaches the true ray direction $\mathbf{r}$, the components of vector $\mathbf{p}_{\mathbf{r}}^T \mathbf{r}_o$ decrease; eventually, the gradient matrix $\mathbf{p}_{\mathbf{r}}$ becomes symmetric and approaches the positive semidefinite Hessian $L_{\mathbf{rr}}$, whose row and column vectors belong to the plane normal to the ray, $\mathbf{p}_{\mathbf{r}} \mathbf{r} = L_{\mathbf{rr}} \mathbf{r} = 0$. This is the solution of the momentum equation H12.

10. Comment. For the "compatible" case, when the slowness and the ray velocity direction match, the following conditions hold,

$$\mathbf{p}_{\mathbf{r}} = \mathbf{p}_{\mathbf{r}}^T \; , \quad \mathbf{p}_{\mathbf{rr}} = \mathbf{p}_{\mathbf{rr}}^T \; , \quad \mathbf{p}_{\mathbf{r}} \cdot \mathbf{r} = 0 \; , \quad \mathbf{p}_{\mathbf{rr}} \cdot \mathbf{r} = -\mathbf{p}_{\mathbf{r}} \quad . \tag{H13}$$

The third-order tensor $\mathbf{p_{rr}}$ is always symmetric wrt the last two indices; for "compatible" case, it becomes supersymmetric with only ten different components: All three indices may be swapped in any order. The Eigenray workflow does not include solving the momentum equation and works with the compatible slowness and ray direction only, after solving the nonlinear set $\mathbf{f} = 0$ (see equation H3) or minimizing the target function in equation D3 of Part I, for the unknown slowness vector and the given ray direction. In this case, equations H7 and H8 simplify to,

$$v_{ray} = \frac{1}{\mathbf{p} \cdot \mathbf{r}} \quad , \quad \frac{\partial v_{ray}}{\partial \mathbf{r}} = -v_{ray}^2 \mathbf{p} \quad , \quad \text{(H14)}$$

$$\frac{\partial^2 v_{ray}}{\partial \mathbf{r}^2} = \frac{2}{v_{ray}} \frac{\partial v_{ray}}{\partial \mathbf{r}} \otimes \frac{\partial v_{ray}}{\partial \mathbf{r}} - v_{ray}^2 \mathbf{p_r} \quad , \quad \text{(H15)}$$

where computing the third-order tensor $\mathbf{p_{rr}}$ becomes unnecessary. We note that matrix $L_{\mathbf{rr}}$ is singular for both approximated ray direction $\mathbf{r}_o$ and its exact value $\mathbf{r}$.

Furthermore, we show that although the resolving matrix of the linearized momentum equation (the Hessian of our first-degree homogeneous Lagrangian wrt the ray direction, $L_{\mathbf{rr}}$) is a positive semidefinite (noninvertible) matrix, the problem is resolvable with an additional scalar constraint, enforcing the ray direction $\mathbf{r}$ to be normalized. We then provide the general constraint equation related to the alternative first-degree homogeneous Lagrangians corresponding to the different flow parameters (e.g., traveltime, arclength, sigma). We assume that the three specified slowness components satisfy the Christoffel equation for the given medium properties.

Solving the nonlinear momentum equation with the first-order homogeneity Lagrangian

Červený (2002a, 2002b) claims that in order to find $\dot{\mathbf{x}}_\zeta \equiv d\mathbf{x}/d\zeta$ using the nonlinear momentum equation $L^\zeta_{\dot{\mathbf{x}}_\zeta}(\mathbf{x},\dot{\mathbf{x}}_\zeta) = \mathbf{p}$, the Lagrangian must be homogeneous of second-degree in $\dot{\mathbf{x}}_\zeta$. In other words, the determinant of the Hessian matrix $L^\zeta_{\dot{\mathbf{x}}_\zeta\dot{\mathbf{x}}_\zeta}$ should not vanish. In this appendix we question this statement.

Recall that in the context of this study, $\dot{\mathbf{x}}_\zeta$ is the ray velocity direction $\mathbf{r}$ (for the flow parameter arclength $\zeta \to s$); for the Lagrangians suggested by Červený (2002a, 2002b), $\dot{\mathbf{x}}_\zeta$ is the ray velocity vector $\dot{\mathbf{x}}_\tau = \mathbf{v}_{\text{ray}}$ (for the flow parameter traveltime $\zeta \to \tau$). Indeed, the condition for the second-degree homogeneity does not hold for our proposed Lagrangian with $\zeta \to s$, which is homogeneous of the first degree in $\mathbf{r} = d\mathbf{x}/ds$, where the Hessian $L^\zeta_{\dot{\mathbf{x}}_\zeta\dot{\mathbf{x}}_\zeta} = L_{\dot{\mathbf{x}}\dot{\mathbf{x}}} \equiv L_{\mathbf{rr}}$ is not invertible, (i.e., $\det L_{\mathbf{rr}} = 0$). However, we show that the requirement for the second-degree homogeneity is inessential. The problem of establishing the components of $\dot{\mathbf{x}}_\zeta$ from the momentum equation (given the slowness vector) is not underdetermined even if $\det L_{\dot{\mathbf{x}}_\zeta\dot{\mathbf{x}}_\zeta} = 0$, since there is always a physical constraint that can be added; in the case of the arclength-related Hamiltonian, the additional constraint is the normalization condition of the ray (or ray velocity) direction, $\mathbf{r} \cdot \mathbf{r} = 1$.

The constraint adjoint to the momentum equation

The physical constraint adjoint to the momentum equation can be arranged in a general form, for an arbitrary flow variable, and it turns to represent the Legendre transform (equation 31). This transform can be formulated as,

$$\dot{\mathbf{x}}_\zeta \cdot \mathbf{p} = L^\zeta\left(\mathbf{x},\dot{\mathbf{x}}_\zeta\right) + H^\zeta\left(\mathbf{x},\mathbf{p}\right) \quad , \tag{H16}$$

where $\zeta$ is an arbitrary flow parameter. The Hamiltonian $H^\zeta$ is constant along the ray, and this constant value, $H^\zeta_{\text{const}}$, should be introduced to equation H16 (instead of the function of the position and slowness),

$$\dot{\mathbf{x}}_\zeta \cdot \mathbf{p} = L^\zeta\left(\mathbf{x},\dot{\mathbf{x}}_\zeta\right) + H^\zeta_{\text{const}} \quad . \tag{H17}$$

For example, the constant values of both, the modified second-degree homogeneity Lagrangian, $L^M$, and its corresponding (eigenvalue) Hamiltonian, $H^\lambda$, are $1/2$ (Červený, 2002a, 2002b), and the constraint becomes, $\dot{\mathbf{x}}_\tau \cdot \mathbf{p} = 1$ (although in this case the constraint is not needed since $\det L^M_{\dot{\mathbf{x}}_\tau \dot{\mathbf{x}}_\tau}$ doesn't vanish). All other Hamiltonians listed in Table 1 of Part I vanish and therefore (although not essential) in the rest of this appendix, we consider the vanishing Hamiltonians. Their corresponding Lagrangians represent the integrand of the traveltime functional, and equation H17 becomes,

$$\dot{\mathbf{x}}_\zeta \cdot \mathbf{p} = L^\zeta = \frac{d\tau}{d\zeta} \quad . \tag{H18}$$

The most common flow parameters are the current traveltime, the arclength and sigma, where equation H18 can be arranged as,

$$d\zeta_n = v_{\text{ray}}^n d\tau \quad , \quad \text{where} \quad \begin{cases} n = 0 & \text{for } \zeta \to \tau \; , \\ n = 1 & \text{for } \zeta \to s \; , \\ n = 2 & \text{for } \zeta \to \sigma \; , \end{cases} \tag{H19}$$

and the following constraint holds due to equation H18 and the identity $\mathbf{v}_{\text{ray}} \cdot \mathbf{p} = 1$,

$$\dot{\mathbf{x}}_{\zeta,n} \cdot \mathbf{p} = v_{\text{ray}}^{-n} \quad , \tag{H20}$$

where $\zeta_n$ is the flow parameter, corresponding to index $n = 0$. Note that for any index value,

$$\dot{\mathbf{x}}_{\zeta,n} = \frac{\dot{\mathbf{x}}_s}{v_{\text{ray}}^{n-1}} = \frac{\mathbf{r}}{v_{\text{ray}}^{n-1}} \quad . \tag{H21}$$

which leads to,

$$\dot{\mathbf{x}}_{\zeta,n} \cdot \dot{\mathbf{x}}_{\zeta,n} = \frac{\mathbf{r}}{v_{\text{ray}}^{n-1}} \cdot \frac{\mathbf{r}}{v_{\text{ray}}^{n-1}} = \frac{1}{v_{\text{ray}}^{2(n-1)}} \quad . \tag{H22}$$

It is suitable to multiply both sides of equation H20 by $v_{\text{ref}}^n$ and both sides of equation H22 by $v_{\text{ref}}^{2(n-1)}$, in order to deal with unitless parameters (as we will use logarithms now). The reference velocity $v_{\text{ref}}$ may be any characteristic value with the units of velocity, for example, the phase velocity (which is known, because the slowness vector is specified), $v_{\text{ref}} = v_{\text{phs}} = 1/\sqrt{\mathbf{p} \cdot \mathbf{p}}$. Combining equations H20 and H22 and applying this normalization, we obtain,

$$\ln\left(\dot{\mathbf{x}}_{\zeta,n} \cdot \mathbf{p} \, v_{\text{phs}}^n\right) = -n \ln \frac{v_{\text{ray}}}{v_{\text{phs}}} \quad , \quad \ln\left(\dot{\mathbf{x}}_{\zeta,n} \cdot \dot{\mathbf{x}}_{\zeta,n} v_{\text{phs}}^{2n-2}\right) = -2(n-1) \ln \frac{v_{\text{ray}}}{v_{\text{phs}}} \quad . \tag{H23}$$

Equation set H23 makes it possible to eliminate the unknown ray velocity magnitude,

Page 73 of 87

$$2(n-1)\ln\left(\dot{\mathbf{x}}_{\zeta,n} \cdot \mathbf{p}\, v_{\text{phs}}^n\right) = n\ln\left(\dot{\mathbf{x}}_{\zeta,n} \cdot \dot{\mathbf{x}}_{\zeta,n} v_{\text{phs}}^{2n-2}\right) \qquad . \tag{H24}$$

Equation H24 can be considered the general constraint equation that accompanies the momentum equation with the first-degree homogeneous Lagrangian; this is a general constraint valid for the flow parameter $\zeta_n$ defined in equation H18. The arguments of logarithmic functions on both sides of equation H24 are unitless.

Consider particular cases. For $n=0$ (flow parameter traveltime), the right-hand side of equation H24 vanishes, and the equation simplifies to,

$$\ln\left(\dot{\mathbf{x}}_\tau \cdot \mathbf{p}\right) = 0 \quad \rightarrow \quad \dot{\mathbf{x}}_\tau \cdot \mathbf{p} = 1 \quad \text{or} \quad \mathbf{v}_{\text{ray}} \cdot \mathbf{p} = 1 \qquad . \tag{H25}$$

For $n=1$ (flow parameter arclength), the left-hand side of equation H24 vanishes, and the equation simplifies to,

$$\ln\left(\mathbf{r} \cdot \mathbf{r}\right) = 0 \quad \rightarrow \quad \mathbf{r} \cdot \mathbf{r} = 1 \qquad . \tag{H26}$$

Finally, for $n=2$ (flow parameter $\sigma$), we obtain,

$$\begin{aligned}\ln\left(\dot{\mathbf{x}}_\sigma \cdot \mathbf{p}\, v_{\text{phs}}^2\right) &= \ln\left(\dot{\mathbf{x}}_\sigma \cdot \dot{\mathbf{x}}_\sigma\, v_{\text{phs}}^2\right), \\ \dot{\mathbf{x}}_\sigma \cdot \mathbf{p} &= \dot{\mathbf{x}}_\sigma \cdot \dot{\mathbf{x}}_\sigma \quad \rightarrow \quad \dot{\mathbf{x}}_\sigma \cdot \left(\dot{\mathbf{x}}_\sigma - \mathbf{p}\right) = 0 ,\end{aligned} \tag{H27}$$

Note that in equation H24, index $n$ can also accept values other than $0,1,2$, including fractional.

<u>Iterative solver for the constrained momentum equation</u>

Next, we develop the iterative solver for the constrained momentum equation, based on the Newton method, for the flow parameters traveltime, $\tau$, and arclength, $s$.

Assume that for a given fixed location $\mathbf{x}$ and a specified slowness vector $\mathbf{p}$, the approximation $\dot{\mathbf{x}}_{\zeta o}$ is known and the correction $\Delta\dot{\mathbf{x}}_{\zeta}$ needs to be found,

$$L^{\zeta}_{\dot{\mathbf{x}}_{\zeta}}\left(\mathbf{x}, \dot{\mathbf{x}}_{\zeta o} + \Delta\dot{\mathbf{x}}_{\zeta}\right) = \mathbf{p} \quad , \tag{H28}$$

where $L^{\zeta}\left(\mathbf{x}, \dot{\mathbf{x}}_{\zeta}\right)$ is a first-degree homogeneous Lagrangian wrt vector $\dot{\mathbf{x}}_{\zeta}$. Linearization of this equation (needed for the iterative numerical solution) leads to,

$$L^{\zeta}_{\dot{\mathbf{x}}_{\zeta}}\left(\mathbf{x}, \dot{\mathbf{x}}_{\zeta o}\right) + L^{\zeta}_{\dot{\mathbf{x}}_{\zeta}\dot{\mathbf{x}}_{\zeta}}\left(\mathbf{x}, \dot{\mathbf{x}}_{\zeta o}\right) \cdot \Delta\dot{\mathbf{x}}_{\zeta} = \mathbf{p} \quad \rightarrow \quad L^{\zeta}_{\dot{\mathbf{x}}_{\zeta}\dot{\mathbf{x}}_{\zeta}}\left(\mathbf{x}, \dot{\mathbf{x}}_{\zeta o}\right) \cdot \Delta\dot{\mathbf{x}}_{\zeta} = \mathbf{p} - L^{\zeta}_{\dot{\mathbf{x}}_{\zeta}}\left(\mathbf{x}, \dot{\mathbf{x}}_{\zeta o}\right) \quad , \tag{H29}$$

which represents an unresolvable linear equation set with a singular matrix, $\det L^{\zeta}_{\dot{\mathbf{x}}_{\zeta}\dot{\mathbf{x}}_{\zeta}}\left(\mathbf{x}, \dot{\mathbf{x}}_{\zeta o}\right) = 0$. We propose two options to overcome this issue using the abovementioned additional constraint. Both options lead to identical results.

Option 1: Replacing one of the three dependent linear equations in set H29 (with the singular matrix) by the constraint.

Option 2: Solving a set of four linearized equations with three variables by least squares. Hence, we don't need to decide, which of the three equations has to be removed. In this particular case, the system of the four equations is not over-determined, and despite the use of the least-squares technique, the solution is exact (the least-squares discrepancy is zero). Option 2 is preferable as, unlike option 1, one does not need to choose, which of the three dependent equations should be replaced by the constraint. Matrix $L_{\mathbf{rr}}$ is positive semidefinite, $L_{\mathbf{rr}}\mathbf{r} = 0$, which means that its three lines (or columns) are coplanar. More specifically, if (unlikely, accidentally) two of the three lines (vectors) are also collinear, then (in the case of using option 1) one of these collinear lines has to be replaced, but not the third line. Hence, using option 2 (the least-squares approach), the inquiry for this rare situation is avoided. We further use this option only.

For the flow parameters traveltime and arclength, the constraints can be presents as,

$$\begin{aligned} \mathbf{v}_{ray} \cdot \mathbf{p} = \left( \mathbf{v}_{ray,o} + \Delta \mathbf{v}_{ray} \right) \cdot \mathbf{p} = 1 & \quad \text{in the case} \quad \zeta \to \tau \quad \text{and} \quad \dot{\mathbf{x}}_\zeta = \dot{\mathbf{x}}_\tau = \mathbf{v}_{ray}, \quad \text{or} \\ \mathbf{r} \cdot \mathbf{r} = \left( \mathbf{r}_o + \Delta \mathbf{r} \right) \cdot \left( \mathbf{r}_o + \Delta \mathbf{r} \right) = 1 & \quad \text{in the case} \quad \zeta \to s \quad \text{and} \quad \dot{\mathbf{x}}_\zeta = \dot{\mathbf{x}} = \mathbf{r}. \end{aligned} \quad (H30)$$

Note that in the case of the flow variable arclength, the initial guess $\mathbf{r}_o$ is normalized, and the correction $\Delta \mathbf{r}$ can be assumed small (and thus, we ignore the square term $\Delta \mathbf{r}^2$). In the case of the flow variable traveltime, one can choose the initial guess for the ray velocity to satisfy the equation,

$$\mathbf{v}_{ray,o} \cdot \mathbf{p} = 1, \quad \text{e.g.,} \quad \mathbf{v}_{ray,o} = \mathbf{v}_{phs} = \frac{\mathbf{p}}{\mathbf{p} \cdot \mathbf{p}}, \quad (H31)$$

where $\mathbf{v}_{phs}$ is the phase velocity vector. Equation set H30 simplifies to,

$$\begin{aligned} \mathbf{r}_o \cdot \Delta \mathbf{r} = 0 & \quad \text{in the case} \quad \zeta = s \quad \text{and} \quad \dot{\mathbf{x}} = \mathbf{r} \\ \Delta \mathbf{v}_{ray} \cdot \mathbf{p} = 0 & \quad \text{in the case} \quad \zeta = \tau \quad \text{and} \quad \dot{\mathbf{x}}_\tau = \mathbf{v}_{ray} \end{aligned}. \quad (H32)$$

Applying the linearized set H29 with the constraint in equation H32, for the flow parameter arclength, the resulting linear system of four equations is written as,

$$\underbrace{\begin{bmatrix} L_{\mathbf{rr}}(\mathbf{x}, \mathbf{r}_o) \\ w \mathbf{r}_o \end{bmatrix}}_{\text{matrix } 4 \times 3} \cdot \Delta \mathbf{r} = \underbrace{\begin{bmatrix} \mathbf{p} - L_{\mathbf{r}} \\ 0 \end{bmatrix}}_{\text{vector of length 4}}. \quad (H33)$$

The constraint is multiplied by a fixed scalar weight, $w$; this is needed to match the units in the resulting relationship. We multiply both sides of this equation by the transpose matrix of the $4 \times 3$ matrix from the left. Recall that the directional Hessian of the Lagrangian is symmetric, $L_{\mathbf{rr}} = L_{\mathbf{rr}}^T$. This leads to,

$$\underbrace{\left(L_{\mathbf{rr}}^2 + w^2\, \mathbf{r}_o \otimes \mathbf{r}_o \right)}_{\text{matrix } 3\times 3} \Delta \mathbf{r} = \underbrace{L_{\mathbf{rr}} \left(\mathbf{p} - L_{\mathbf{r}}\right)}_{\text{vector of length 3}} . \qquad (H34)$$

Note that the product $L_{\mathbf{rr}} \mathbf{r}_o$ vanishes, which means that the three lines of matrix $L_{\mathbf{rr}}$ are coplanar and belong to the plane normal to $\mathbf{r}_o$. This, in turn, implies that the matrix on the left side of equation H34 is never singular (although each of its two counterparts in brackets is singular). These two items should have the same units. The Hessian $L_{\mathbf{rr}}$ has the units of slowness, while the ray direction $\mathbf{r}_o$ is unitless. Thus, the weight has the unit of slowness as well. The most reasonable assertion is, $w^2 = p^2 = \mathbf{p} \cdot \mathbf{p}$.

A similar approach can be applied for the momentum equation with the time-related, unmodified, first-degree homogeneity Lagrangian, $L^U(\mathbf{x}, \dot{\mathbf{x}}_\tau)$,

$$\underbrace{\begin{bmatrix} L^U_{\dot{\mathbf{x}}_\tau \dot{\mathbf{x}}_\tau}(\mathbf{x}, \dot{\mathbf{x}}_{\tau,o}) \\ w\mathbf{p} \end{bmatrix}}_{\text{matrix } 4\times 3} \cdot \Delta \dot{\mathbf{x}}_\tau = \underbrace{\begin{bmatrix} \mathbf{p} - L^U_{\dot{\mathbf{x}}_\tau} \\ 0 \end{bmatrix}}_{\text{vector of length 4}} , \qquad (H35)$$

which leads to,

$$\underbrace{\left[\left(L^U_{\dot{\mathbf{x}}_\tau\dot{\mathbf{x}}_\tau}\right)^2 + w^2\,\mathbf{p}\otimes\mathbf{p}\right]}_{\text{matrix 3}\times\text{3}}\Delta\dot{\mathbf{x}}_\tau = \underbrace{L^U_{\dot{\mathbf{x}}_\tau\dot{\mathbf{x}}_\tau}\left(\mathbf{p}-L^U_{\dot{\mathbf{x}}_\tau}\right)}_{\text{vector of length 3}} \quad , \tag{H36}$$

(with the weight, $w^2 = p^2$, as well). Note that for the tensor products in equations H34 and H36, the following identities hold,

$$\left(\mathbf{r}_o \otimes \mathbf{r}_o\right)\Delta\mathbf{r} = \left(\mathbf{r}_o \cdot \Delta\mathbf{r}\right)\mathbf{r}_o \quad \text{and} \quad \left(\mathbf{p}\otimes\mathbf{p}\right)\Delta\dot{\mathbf{x}}_\tau = \left(\mathbf{p}\cdot\Delta\dot{\mathbf{x}}_\tau\right)\mathbf{p} \quad . \tag{H37}$$

Solutions of linear sets H34 and H36 are independent of the assigned weights, $w$. In these solutions, the scalar products $\mathbf{r}_o \cdot \Delta\mathbf{r}$ and $\mathbf{p}\cdot\Delta\dot{\mathbf{x}}_\tau$ vanish, respectively. This is, of course, valid only for the singular Hessian matrices, $L_{\mathbf{rr}}$ and $L^U_{\dot{\mathbf{x}}_\tau\dot{\mathbf{x}}_\tau}$.

Enforcing the constraint

In the case of the flow parameter arclength and our suggested Lagrangian, we solve equation H34 for the correction vector $\Delta\mathbf{r}$, and we then compute the updated normalized ray direction,

$$\mathbf{r}_{(i+1)} = \frac{\mathbf{r}_{(i)} + \Delta\mathbf{r}}{\left|\mathbf{r}_{(i)} + \Delta\mathbf{r}\right|} \quad , \tag{H38}$$

where $\mathbf{r}_{(i)}$ is the ray direction at or at the previous iteration.

In case of the unmodified, first-degree homogeneity, traveltime-related Lagrangian suggested by Červený (2002a, 2002b), the analog to the normalization update equation H38 is,

$$\dot{\mathbf{x}}_{\tau,(i+1)} = \frac{\dot{\mathbf{x}}_{\tau,(i)} + \Delta\dot{\mathbf{x}}_\tau}{\left[\dot{\mathbf{x}}_{\tau,(i)} + \Delta\dot{\mathbf{x}}_\tau\right]\cdot\mathbf{p}} \quad . \tag{H39}$$

This update holds the constraint and preserves the direction of the non-normalized updated ray velocity vector in the numerator on the left-hand side. However, a preferable normalization (which may lead probably to a faster convergence), is the minimum change of the vector that holds the constraint,

$$\dot{\mathbf{x}}_{\tau,(i+1)} = \dot{\mathbf{x}}_{\tau,(i)} + \Delta\dot{\mathbf{x}}_\tau + \Delta\dot{\mathbf{x}}_\tau \quad , \tag{H40}$$

where the additional (normalizing) correction $\Delta\dot{\mathbf{x}}_\tau$ satisfies the following conditions,

$$\left[\dot{\mathbf{x}}_{\tau,(i)} + \Delta\dot{\mathbf{x}}_\tau + \Delta\dot{\mathbf{x}}_\tau\right] \cdot \mathbf{p} = 1 \quad \text{and} \quad \Delta\dot{\mathbf{x}}_\tau \cdot \Delta\dot{\mathbf{x}}_\tau \to \min \quad . \tag{H41}$$

The result of this constrained optimization reads,

$$\Delta\dot{\mathbf{x}}_\tau = -\frac{\left[\dot{\mathbf{x}}_{\tau,(i)} + \Delta\dot{\mathbf{x}}_\tau\right] \cdot \mathbf{p} - 1}{\mathbf{p} \cdot \mathbf{p}} \mathbf{p} \quad . \tag{H42}$$

Here $\Delta\dot{\mathbf{x}}_\tau$ is the non-normalized correction resulting from the iterative method (equation H36), and $\Delta\dot{\mathbf{x}}_\tau$ is the additional normalizing update.

Numerical example

In order to demonstrate this functionality, we consider a numerical example using the triclinic medium studied in appendix F. Assume the following slowness components (equation F5),

$$\mathbf{p} = \begin{bmatrix} 0.12893238 & 0.25360427 & 0.14179428 \end{bmatrix} \text{s/km} \quad , \tag{H43}$$

and find the ray velocity direction $\mathbf{r}$. Let the initial guess $\mathbf{r}_o$ coincide with the slowness direction,

$$\mathbf{r}_o = \frac{\mathbf{p}}{\sqrt{\mathbf{p} \cdot \mathbf{p}}} = [0.40560752 \quad 0.79781199 \quad 0.44607015] \quad . \tag{H44}$$

For the given slowness vector $\mathbf{p}$ and the approximate ray direction $\mathbf{r}_o$, we compute the ray velocity and its directional gradient and Hessian (km/s),

$$v_{\text{ray}} = 3.1458940 \;,\qquad \nabla_{\mathbf{r}} v_{\text{ray}} = \begin{bmatrix} -4.6525193 \cdot 10^{-2} \\ +7.1416298 \cdot 10^{-3} \\ +2.9531909 \cdot 10^{-2} \end{bmatrix},$$

$$\nabla_{\mathbf{r}} \nabla_{\mathbf{r}} v_{\text{ray}} = \begin{bmatrix} +2.9310654 & -1.1579785 & -0.48980642 \\ -1.1579785 & +1.2208760 & -1.1466513 \\ -0.48980642 & -1.1466513 & +2.4300002 \end{bmatrix} . \tag{H45}$$

Next, applying equations H10 and H11, we compute the directional gradient $L_{\mathbf{r}}$ and Hessian $L_{\mathbf{rr}}$ of the proposed Lagrangian (s/km),

$$L_{\mathbf{r}} = \begin{bmatrix} +0.13363348 \\ +0.25288265 \\ +0.13881025 \end{bmatrix} \neq \mathbf{p} \;,\qquad \left| \mathbf{p}_{\mathbf{r}}^T \mathbf{r}_{(1)} \right| = 6.088 \cdot 10^{-3} \;,$$

$$L_{\mathbf{rr}} = \begin{bmatrix} -2.6636083 \cdot 10^{-2} & +1.7579873 \cdot 10^{-2} & -7.2222754 \cdot 10^{-3} \\ +1.7579873 \cdot 10^{-2} & -8.9646337 \cdot 10^{-3} & +4.8341862 \cdot 10^{-5} \\ -7.2222754 \cdot 10^{-3} & +4.8341862 \cdot 10^{-5} & +6.4806945 \cdot 10^{-3} \end{bmatrix} . \tag{H46}$$

As expected, matrix $L_{\mathbf{rr}}$ is singular (its determinant is zero). Next, we apply the linearized equation set H34, where the difference $\mathbf{p} - L_{\mathbf{r}}$ is the discrepancy of the approximation, that tends to zero as the ray velocity direction is being refined. This matrix in this equation set is invertible. We solve it for the correction vector $\Delta \mathbf{r}$, and we then compute the updated normalized ray direction with equation H38, which leads to,

$$\mathbf{r}_{(1)} = [0.31742134 \quad 0.55308295 \quad 0.77028757] \quad , \tag{H47}$$

where the subscript in brackets indicates the iteration number. We continue the iterative procedure, and after seven iterations, we obtain the "exact" ray velocity direction (see equation F3) with 13 digits correct, while the expression in brackets on the right-hand side of the matrix equation set H34 vanishes. The resulting values are,

$$v_{ray} = 3.4489725 \quad , \qquad \nabla_{\mathbf{r}} v_{ray} = \begin{bmatrix} -0.76113381 \\ -0.94734363 \\ +0.96210957 \end{bmatrix} \quad ,$$

$$\nabla_{\mathbf{r}} \nabla_{\mathbf{r}} v_{ray} = \begin{bmatrix} +3.4931631 & +0.73180370 & -0.59950123 \\ +0.73180370 & +3.7665636 & -1.9225502 \\ -0.59950123 & -1.9225502 & +0.42410003 \end{bmatrix} \quad , \tag{H48}$$

$$L_{\mathbf{r}} = \begin{bmatrix} +0.12893238 \\ +0.25360427 \\ +0.14179428 \end{bmatrix} = \mathbf{p} \quad , \qquad \left| \mathbf{p}_{\mathbf{r}}^T \mathbf{r}_{(7)} \right| = 4.418 \cdot 10^{-17} \quad ,$$

$$L_{\mathbf{rr}} = \begin{bmatrix} +3.8643656 \cdot 10^{-2} & -9.1071062 \cdot 10^{-3} & -4.1561396 \cdot 10^{-3} \\ -9.1071062 \cdot 10^{-3} & +8.2395475 \cdot 10^{-3} & -3.7809072 \cdot 10^{-3} \\ -4.1561396 \cdot 10^{-3} & -3.7809072 \cdot 10^{-3} & +4.1660411 \cdot 10^{-3} \end{bmatrix} \quad , \tag{H49}$$

$$\mathbf{r}_{(7)} = [0.224 \quad 0.600 \quad 0.768] \quad \text{and} \quad L_{\mathbf{r}} = \mathbf{p} \quad . \tag{H50}$$

We make sure that the resulting directional gradient of the ray velocity is normal to the ray direction, $\nabla_\mathbf{r} v_{\text{ray}} \cdot \mathbf{r} = 0$, and also $L_{\mathbf{rr}}\mathbf{r} = 0$ (equations H48 – H50); this is not so for the initial guess (equations H44 – H46).

Closing remark

We emphasize that this numerical example is provided only to demonstrate that the ray velocity direction can be obtained by an iterative linearization of the momentum equation, $L_\mathbf{r}(\mathbf{x},\mathbf{r}) = \mathbf{p}$, even for the first-degree homogeneous Lagrangian, given the medium properties at the specified location $\mathbf{x}$ and the slowness vector $\mathbf{p}$. The singularity of the Hessian matrix $L_{\mathbf{rr}}$ (this matrix appears due to the linearization) is removed due to the additional constraint asserting that the ray direction $\mathbf{r}$ is normalized. We recall that the Eigenray workflow does not involve solving the momentum equation.

## LIST OF TABLES

Table 1. Elastic properties of the triclinic model.

## LIST OF FIGURES

Figure 1. Hamiltonian and Lagrangian approaches for ray theory and particle mechanics.

Table 1. Elastic properties of the triclinic model.

| Component, km$^2$/s$^2$ | | | Relative gradient, km$^{-1}$ | | |
|---|---|---|---|---|---|
| # | $C_{ij}$ | value | $x_1$ | $x_2$ | $x_3$ |
| 1 | $C_{11}$ | +45.597 | +0.4647 | +0.1201 | +0.2126 |
| 2 | $C_{12}$ | +0.935 | +0.2993 | +0.0965 | +0.1592 |
| 3 | $C_{13}$ | +0.895 | +0.1517 | –0.0708 | +0.2103 |
| 4 | $C_{14}$ | +0.103 | +0.0062 | +0.3696 | –0.0399 |
| 5 | $C_{15}$ | +0.074 | –0.0308 | –0.0730 | +0.0514 |
| 6 | $C_{16}$ | +0.070 | +0.3953 | +0.2842 | +0.3468 |
| 7 | $C_{22}$ | +14.442 | –0.0786 | +0.0767 | +0.3028 |
| 8 | $C_{23}$ | +0.887 | +0.1143 | +0.1624 | +0.3564 |
| 9 | $C_{24}$ | –0.083 | +0.1668 | –0.0605 | +0.4459 |
| 10 | $C_{25}$ | +0.018 | +0.3221 | +0.3742 | +0.3664 |
| 11 | $C_{26}$ | –0.059 | +0.4380 | +0.1161 | +0.4072 |
| 12 | $C_{33}$ | +44.303 | +0.1007 | +0.2677 | +0.0293 |
| 13 | $C_{34}$ | –0.049 | +0.1362 | +0.1640 | +0.1998 |
| 14 | $C_{35}$ | –0.040 | –0.0412 | +0.2176 | +0.0890 |
| 15 | $C_{36}$ | –0.026 | +0.2618 | +0.2117 | +0.4530 |
| 16 | $C_{44}$ | +0.459 | –0.0788 | +0.4569 | +0.0150 |
| 17 | $C_{45}$ | +0.090 | +0.0019 | +0.3758 | +0.0294 |
| 18 | $C_{46}$ | +0.052 | +0.1136 | +0.2295 | +0.0055 |
| 19 | $C_{55}$ | +0.374 | +0.1294 | +0.1885 | +0.4181 |
| 20 | $C_{56}$ | +0.101 | +0.1619 | +0.4534 | +0.2185 |
| 21 | $C_{66}$ | +0.450 | +0.1679 | +0.3936 | +0.4576 |

**Ray Theory Hamiltonian** (initial value problem)

$$H^\zeta(\mathbf{x},\mathbf{p}) = \sum_{j=1}^{3} p_j \dot{x}_{\zeta,j}(\mathbf{x},\mathbf{p}) - L^\zeta\left[\mathbf{x},\dot{\mathbf{x}}_\zeta(\mathbf{x},\mathbf{p})\right] \quad \text{(Legendre trans.)},$$

$\mathbf{x} = \mathbf{x}(\zeta)$, $\mathbf{p} = \mathbf{p}(\zeta)$, $dH^\zeta = 0$ along the ray $\Rightarrow$

$$\dot{x}_{\zeta,j} = \frac{\partial H^\zeta(\mathbf{x},\mathbf{p})}{\partial p_j}, \quad \dot{p}_j = -\frac{\partial H^\zeta(\mathbf{x},\mathbf{p})}{\partial x_j}, \quad j = \{1,2,3\}$$

6 first-order ODE;  particle mechanics: $H^\zeta = E_k + E_p$

**Ray Theory Lagrangian** (boundary value problem)

Principle of least (stationary) action

$$L^\zeta(\mathbf{x},\dot{\mathbf{x}}_\zeta) = \sum_{j=1}^{3} p_j(\mathbf{x},\dot{\mathbf{x}}_\zeta)\dot{x}_{\zeta,j} - H^\zeta\left[\mathbf{x},\mathbf{p}(\mathbf{x},\dot{\mathbf{x}}_\zeta)\right] \quad \text{(Legendre trans.)},$$

$\mathbf{x} = \mathbf{x}(\zeta)$, $\dot{\mathbf{x}}_\zeta = \dot{\mathbf{x}}_\zeta(\zeta)$, $\left(dL^\zeta = 0 \text{ along the ray for } \zeta = \tau\right)$

$$p_j = \frac{\partial L^\zeta(\mathbf{x},\dot{\mathbf{x}}_\zeta)}{\partial \dot{x}_{\zeta,j}}, \quad \dot{p}_j = \frac{\partial L^\zeta(\mathbf{x},\dot{\mathbf{x}})}{\partial x_j}, \quad j = \{1,2,3\}$$

3 second-order ODE; particle mechanics: $L^\zeta = E_k - E_p$

Action: $J = \int_a^b L_\zeta(\mathbf{x},\dot{\mathbf{x}}_\zeta) d\zeta \to \min$,  $(\delta J = 0) \Rightarrow$

$$\frac{d}{d\zeta}\frac{\partial L}{\partial \dot{\mathbf{x}}_\zeta} = \frac{\partial L}{\partial \mathbf{x}} \quad \text{(Euler-Lagrange)}$$

particle mechanics: $\sum_{j=1}^{3} p_j \dot{x}_{\zeta,j} = 2E_k$

---

$\mathbf{x}$: spatial coordinates , $\mathbf{p}$: slowness vector , $\dot{\mathbf{x}}_\zeta = \dfrac{d\mathbf{x}}{d\zeta}$

for $\zeta = \tau$ (time)   $\dot{\mathbf{x}}_\tau = \mathbf{v}_{ray}$ (ray velocity) and $\mathbf{p}\cdot\mathbf{v}_{ray} = 1$

for $\zeta = s$ (arclength) $\dot{\mathbf{x}} = \mathbf{r}$   (ray direction) and $\mathbf{p}\cdot\mathbf{r} = \dfrac{1}{v_{ray}}$

Figure 1. Hamiltonian and Lagrangian for ray theory and particle mechanics.